 \documentclass{iopart}

\usepackage{iopams} 
\usepackage{isolatin1}
\usepackage{graphicx}
\usepackage[german]{varioref}
\usepackage[dvips]{epsfig}
\usepackage{graphics}
\usepackage[dvips]{color}
\usepackage{color}
\usepackage{colordvi}
\usepackage{dsfont}
\usepackage{psfrag}
\begin{document}
\newcommand{\be}{\begin{eqnarray}}
\newcommand{\eps}{\epsilon}
\newcommand{\ee}{\end{eqnarray}}
\def\p#1#2{|#1\rangle \langle #2|}
\def\ket#1{|#1\rangle}
\def\bra#1{\langle #1|}
\def\refeq#1{(\ref{#1})}
\def\tb#1{{\overline{{\underline{ #1}}}}}
\def\im{\mbox{Im}}
\def\re{\mbox{Re}}
\def\nn{\nonumber}
\def\t{\mbox{tr}}
\def\sgn{\mbox{sgn}}
\def\Li{\mbox{Li}}
\def\P{\mbox{P}}
\def\d{\mbox d}
\def\i{\int_{-\infty}^{\infty}}
\def\ip{\int_{0}^{\infty}}
\def\mi{\int_{-\infty}^{0}}
\def\A{\mathfrak A}
\def\AA{{\overline{{\mathfrak{A}}}}}
\def\a{\mathfrak a}
\def\aa{{\overline{{\mathfrak{a}}}}}
\def\B{\mathfrak B}
\def\BB{{\overline{{\mathfrak{B}}}}}
\def\b{\mathfrak b}
\def\bb{{\overline{{\mathfrak{b}}}}}
\def\R{\mathcal R}
\def\dm{\mathfrak d}
\def\dd{{\overline{{\mathfrak{d}}}}}
\def\D{\mathfrak D}
\def\DD{{\overline{{\mathfrak{D}}}}}
\def\c{\mathfrak c}
\def\cc{{\overline{{\mathfrak{c}}}}}
\def\C{\mathfrak C}
\def\CC{{\overline{{\mathfrak{C}}}}}
\def\O{\mathcal O}
\def\F{\mathcal F_k}
\def\N{\mathcal N}
\def\I{\mathcal I}
\def\S{\mathcal S}
\def\P{\mathcal P}

\def\G{\Gamma}
\def\L{\Lambda}
\def\la{\lambda}
\def\g{\gamma}
\def\al{\alpha}
\def\s{\sigma}
\def\e{\epsilon}
\def\te{{\rm e}}
\def\max{{\rm max}}
\def\str{{\rm str}}
\def\tC{\text C}
\def\Fo{\mathcal{F}_{1,k}}
\def\Ft{\mathcal{F}_{2,k}}
\def\vs{\varsigma}
\def\l{\left}
\def\r{\right}
\def\up{\uparrow}
\def\down{\downarrow}
\def\u{\underline}
\def\ov{\overline}
\title[Periodic and hard wall fermions and bosons]{Bethe 
Ansatz study of one-dimensional Bose and Fermi gases with periodic
and hard wall boundary conditions
}
\author{N. Oelkers\footnote{Email: oelkers@maths.anu.edu.au}, M. T. Batchelor, 
M. Bortz and X.-W. Guan
}

\address{Department of Theoretical Physics, Research School of Physical Sciences \& Engineering and
Mathematical Sciences Institute, 
The Australian National University,  Canberra ACT 0200, Australia}

\begin{abstract}
We extend the exact periodic Bethe Ansatz solution for one-dimensional 
bosons and fermions with $\delta$-interaction and arbitrary internal degrees 
of freedom to the case of hard wall boundary conditions.
We give an analysis of the ground state properties of fermionic systems with two internal 
degrees of freedom, including expansions of the ground state energy in the weak and 
strong coupling limits.
\end{abstract}




\section{Introduction}
One-dimensional quantum gases with two-particle $\delta$-interaction have long been of 
fascination.
The most simple model of $\delta$-interacting spinless bosons in a periodic box was solved in 
terms of the Bethe Ansatz by Lieb and Liniger \cite{LL}.
This quantum mechanical model is not only one of the oldest integrable models
after the Heisenberg spin chain, but arguably also one of the most important test beds for
exploring new ideas and methods, e.g., the Thermodynamic 
Bethe Ansatz \cite{TBApioneer} was pioneered for this model. 
Earlier Girardeau \cite{Girardeau} 
discussed a mapping from strongly repulsive interacting bosons to fermions, 
corresponding to the limit $c\!\to\!\infty$, where $c$ features as the arbitrary interaction 
strength in the Lieb-Liniger model.
Later in seminal work McGuire \cite{McGuire} discussed $\delta$-interaction 
particles via an optical analogue.
Gaudin \cite{Gaud} and Yang \cite{Yang} then considered spin-$\case12$ fermions 
with periodic boundary conditions, the first model with internal states.
Sutherland \cite{Sutherland} applied the nested Bethe Ansatz, which allowed the treatment of periodic quantum 
gases with arbitrary spin by repeated application of the Bethe Ansatz, reducing the number of internal states 
in each step.
Gaudin \cite{Gaudin} solved the model of spinless bosons with hard wall boundary conditions.

The special form of $\delta$-interaction is at the heart of the integrability of the quantum gases.
Models tweaking the type of interaction have been considered, but so far are less 
prominent \cite{unusual-Generalizations}.
Non-integrable models, like the harmonically trapped gas with tunable interaction strength, 
have to be treated by approximate methods and simulations.
One integrable model, the mixture of fermionic and bosonic particles, solved by Lai and Yang \cite{Lai} 
has long been dormant in the literature, but is enjoying new interest \cite{mixture-papers} in light
of the novel recent experiments on low-dimensional quantum gases.
Apart from isolated work \cite{other-mixture-model} this is the only integrable approach to 
mixtures of bosons and fermions.

Generalizations of the original models and the evaluation of their physical properties
have been discussed extensively in several 
books \cite{Gaudinbook,Mabook,Korepin,Many-Particle-Physics,Takahashi,Sutherlandbook}
and review articles \cite{Schlottman}, mostly concentrating on periodic 
spinless bosons and periodic spin-$\frac12$ fermions.
Of direct relevance here are a series of papers examining 
periodic $su(2)$ \cite{periodicsu2bosons} and $su(3)$ \cite{periodicsu3bosons} bosons 
as well as hard wall $su(2)$ bosons \cite{hwsu2bosons} via the Bethe Ansatz.
Hard wall fermions with two internal states have been treated 
numerically for up to three particles \cite{HWsu2fermions-numerics}, 
but no Bethe Ansatz solution was given.
A technically involved model in which the hard walls are of finite height, 
resulting in tunnelling outside the box, has also been considered \cite{su2fermions-finite-well}.
However, only the case of one particle outside the box could be treated and no physical 
properties where evaluated due to the complicated nature of the solution.
In 1985 {Woynarovich}~\cite{Woynarovich}
discussed a hard wall spin-$\frac12$ fermion model with an additional internal potential. 
The Bethe Ansatz solution
he obtained differs from the standard model by one additional term
The $su(N)$ pure fermion model with $\delta$-potentials of variable strength at the boundaries, acting like transparent walls,
was also solved~\cite{Fujimoto} by extending this idea.
This system reduces to the hard wall model considered here for $|\gamma_0|,|\gamma_L| \to \infty$,
where $\gamma_0$ and $\gamma_L$ act as boundary potentials.
Finally we note that the Hubbard model with open ends was also solved exactly \cite{Schulz}.
In an appropriate continuum limit \cite{KluemperHubbardbook} the resulting integral equations 
reduce to the integral equations for the ground state energy of a hard wall fermion gas with spin-$\frac12$.

The content and arrangement of this paper is as follows.
In Section \ref{sec:model} we discuss the most general one-dimensional integrable $\delta$-interaction 
model and its exact solution, both for periodic and hard wall boundary conditions.
The treatment for periodic boundary conditions serves to highlight the key differences in the 
solution for hard wall boundary conditions.
In Sections \ref{sec:weak} and \ref{sec:hard} we give an approximate treatment of the 
Bethe Ansatz solution for the special case of the spin-$\frac12$ fermionic gas with 
periodic and hard wall boundary conditions
via methods developed in References~\cite{ours-hw-bosons,ours-TG,ours-fermions}.
The ground state energy in the thermodynamic limit
is considered via integral equations in Section \ref{sec:NLIE}.
%
%
Concluding remarks are given in Section \ref{sec:outlook}.
The detailed derivation of the coordinate Bethe Ansatz for the first level and
the Algebraic Bethe Ansatz for the nested structure for bosons
and fermions with arbitrary spin and either periodic or hard wall boundary conditions 
is given in the Appendices.

\section{Exact solution}
\label{sec:model}

As already remarked, special cases of the exact solutions presented here
have been examined for more than 40 years. 
%
The one-dimensional $\delta$-interaction quantum gas with $N$ particles is described by 
wave functions $\Psi(x_1,\ldots,x_N,\sigma_1,\ldots,\sigma_N)$ which are eigenfunctions
of the Hamiltonian 
$\mathbb{H}=\mathcal{H} \oplus \textrm{Id}_{V_1} \oplus \cdots \oplus \textrm{Id}_{V_N}$, 
where 
\be
\mathcal{H}=-\sum_{i=1}^N \frac{\partial^2}{\partial x_i^2}
+ 2 c \sum_{i<j} \delta({x_i - x_j}).
\ee
Here the space coordinates, $x_i\in[0,L], i=1,\ldots,N$, occur only in the space part $\mathcal{H}$, 
which operates on the standard space of square integrable functions
$L^2[0,L]$, as usually used in quantum mechanics.
The spin coordinates $\sigma_i$ take values in $\sigma_i \!\in\! \{1,\ldots,N'\}$ 
and $\sigma_i$ acts trivially on the $N'$-dimensional vector space $V_i$.

Here we initially solve the problem for an {\sl abstract} model with $N'$ internal states.
The name $su(N')$ model results from the symmetry of the underlying integrable model 
of the $su(N')$ chain, but one can make the case that this quantum gas Hamiltonian 
also possesses $su(N')$ symmetry itself.
For physical application these internal states are usually interpreted as various distinguishing 
physical properties, for instance spin or isospin/hyperfine states.
E.g., the su(2) model has two internal states, which can be interpreted as $\uparrow$ and $\downarrow$.
The coordinates $\sigma_i$ only contribute to the symmetry and do not enter in the functional
form of the solution.

The free parameter $c$ acts as an attractive (repulsive) interaction strength for $c\!<\!0\,\, (c\!>\!0)$
between two particle space coordinates $x_i$ and $x_j$.
The eigenfunctions $\Psi$ of Hamiltonian $\mathbb{H}$ and their corresponding eigenvalues 
have to satisfy three additional properties to be a solution for the quantum gas:
\begin{enumerate}
\item Symmetry, with $\Psi \to \pm \Psi$ under any `particle exchange' 
$(x_i,\sigma_i)\leftrightarrow (x_j,\sigma_j),\ \ i,j=1,\ldots,N$. 
Here the upper (lower) sign denotes a gas of bosons (fermions).
\item  Continuity of $\Psi$ in the interval $[0,L]$ in each space coordinate $x_i$.
\item Periodic boundary conditions 
$\Psi(\ldots \,x_i=0 \,\ldots)=\Psi(\ldots \,x_i=L\, \ldots)$ or hard wall boundary conditions 
$\Psi(\ldots\,x=0\, \ldots)=\Psi(\ldots\,x_i=L\,\ldots)=0$, $i=1,\ldots,N$, for a given box of size $L$.
\end{enumerate}
This model is known to be Yang-Baxter-integrable and its Bethe Ansatz solution is described in 
detail in \ref{appa} for the given boundary conditions.
Each energy eigenvalue is described by $N'$ sets of complex parameters, where all
these sets together form the coupled algebraic equations given below -- 
the Bethe equations.
For all models of this type the energy eigenvalue is determined solely by the quasi-momenta 
$\{k_1,\ldots,k_N\}$, which are the roots of the 1st level or 1st set, with
\be
E=\sum_{i=1}^N k_i^2.
\ee
To determine the allowed values for $\{k_i\}$, i.e., the energy eigenvalues, it is necessary to
find a solution that satisfies the complete set of equations. Consider first the periodic case.
These equations are
\be
\fl
\qquad
\mathrm{e}^{\mathrm{i}k_i L} =
\prod_{j=1}^{M_1} \frac{k_i - \Lambda^{(1)}_j + \frac12 \mathrm{i} c }{  k_i - \Lambda^{(1)}_j - \frac12 \mathrm{i} c }
\qquad
\mathrm{e}^{\mathrm{i}k_i L} =
\prod_{j\neq i}^N \frac{k_i - k_j + \mathrm{i}c  }{  k_i - k_j - \mathrm{i}{c}  }
\prod_{j=1}^{M_1} \frac{k_i - \Lambda^{(1)}_j -\frac12 \mathrm{i} c }{  k_i - \Lambda_j^{(1)} + \frac12 \mathrm{i} c  }
\ee
where the formula on the left (right) is for fermions (bosons). 
Both fermions and bosons are subject to the same higher level Bethe equations, namely
\be
\fl 
\qquad
\prod_{j =1}^{M_{l-1}} 
\frac{\Lambda^{(l)}_i - \Lambda^{(l-1)}_j + \frac12 \mathrm{i} c  }
{\Lambda^{(l)}_i - \Lambda^{(l - 1)}_j - \frac12 \mathrm{i} c  }
=
\prod_{j\neq i}^{M_l} 
\frac{\Lambda^{(l)}_i - \Lambda^{(l)}_j + \mathrm{i}{c} }{\Lambda^{(l)}_i - \Lambda^{(l)}_j - \mathrm{i}c}
%
\prod_{j=1}^{M_{l+1}}
 \frac{\Lambda^{(l)}_i - \Lambda^{(l+1)}_j  - \frac12 \mathrm{i} c  }{
\Lambda^{(l)}_i - \Lambda^{(l+1)}_j + \frac12 \mathrm{i} c }
\label{eq:suN-periodic-BAE}
\ee
The Bethe equations for the hard wall case are
\be
\fl
\qquad
\mathrm{e}^{\mathrm{i}2 k_i L} =
\prod_{j =1}^{M_1} 
\frac{k_i - \Lambda^{(1)}_j + \frac12 \mathrm{i} c}{k_i - \Lambda^{(1)}_j - \frac12 \mathrm{i} c  }
\ \frac{k_i + \Lambda^{(1)}_j + \frac12 \mathrm{i} c   }{  k_i + \Lambda^{(1)}_j - \frac12 \mathrm{i} c  }
\nonumber
 \\
\fl
\qquad
\mathrm{e}^{\mathrm{i}2 k_i L} =
\prod_{j\neq i}^N 
\frac{k_i - k_j + \mathrm{i}c  }{  k_i - k_j - \mathrm{i}{c}  }
\overbrace{
\frac{k_i + k_j + \mathrm{i}c  }{  k_i + k_j - \mathrm{i}{c}  }
}^A
\prod_{j=1}^{M_1} 
\frac{k_i - \Lambda^{(1)}_j - \frac12 \mathrm{i} c  }{  k_i - \Lambda^{(1)}_j + \frac12 \mathrm{i} c  }
\overbrace{
\frac{k_i + \Lambda^{(1)}_j - \frac12 \mathrm{i} c   }{  k_i + \Lambda^{(1)}_j + \frac12 \mathrm{i} c  }
}^A
\label{eq:HW-1st-level-BAE-suN}
\ee
where now the formula on the 1st (2nd) line is for fermions (bosons). 
Both bosonic and fermionic systems are subject to the same higher level 
Bethe equations, 
\be
\fl
\quad
\prod_{j =1}^{M_{l-1}} 
\frac{\Lambda^{(l)}_i - \Lambda^{(l-1)}_j + \frac12 \mathrm{i} c }
{\Lambda^{(l)}_i - \Lambda^{(l-1)}_j -\frac12 \mathrm{i} c }
\, \frac{\Lambda^{(l)}_i + \Lambda^{(l-1)}_j + \frac12 \mathrm{i} c }
{\Lambda^{(l)}_i + \Lambda^{(l-1)}_j - \frac12 \mathrm{i} c }
=
\prod_{j\neq i}^{M_l} 
\frac{\Lambda^{(l)}_i - \Lambda^{(l)}_j + \mathrm{i}{c} }{\Lambda^{(l)}_i - \Lambda^{(l)}_j - \mathrm{i}c}
\frac{\Lambda^{(l)}_i + \Lambda^{(l)}_j + \mathrm{i}{c} }{\Lambda^{(l)}_i + \Lambda^{(l)}_j - \mathrm{i}c}
\nonumber\\
\times
\prod_{j=1}^{M_{l+1}} 
\frac{\Lambda^{(l)}_i\!-\!\Lambda^{(l+1)}_j - \frac12 \mathrm{i} c }
{\Lambda^{(l)}_i - \Lambda^{(l+1)}_j + \frac12 \mathrm{i} c}
\,
\frac{\Lambda^{(l)}_i + \Lambda^{(l +1)}_j - \frac12 \mathrm{i} c }
{\Lambda^{(l)}_i + \Lambda^{(l + 1)}_j + \frac12 \mathrm{i} c}.
\label{eq:suN-HW-higher-levels}
\ee

The above notation for the periodic and the hard wall boundary conditions denotes 
a large set of equations, in the $l-th$ level root set
$\{\Lambda^{(l)}_i\}$ there are $i=1,\ldots,M_l$ roots and as many equations. 
To make the notation compact we have set $\Lambda^{(0)}_i\equiv k_i$ and $M_0\equiv N$ 
as well as $M_{N'}\equiv0$, i.e., the product $\prod_{j=1}^{M_{N'}}=1$ is unity.
Here the numbers of roots in each set is restricted by 
$\frac12 {N}\geq\! M_1 \!\geq\! ...\!\geq\! M_{N'\textendash1} \geq 0$.
The connection between the number of spin coordinates $\sigma_i,\ldots,\sigma_N$
in the internal states $1,\ldots,N'$ on the one side, and the quantum numbers $M_i$ 
on the other, is given by $M_{N'-1}=N_{N'}$, $M_{N'-2}=N_{N'-1} + N_{N'}$, $\ldots$, 
$M_{1}=N_{2}+ \dots + N_{N'}$, $N=N_1 + N_{2}+ \ldots +N_{N'}$.
Here we use a convention to label the internal state, in which the most $\sigma_i$
are in, as $1$, with $N_1$ the number of particles in $1$. 
Similarly the internal state in which the second-most number of $\sigma_i$ are in
is named $2$ with $N_2$ particles in it, until we reach $N'$, the internal state with the 
least number of $\sigma_i$ in it, namely $N_{N'}$.
%
%
This order is no restriction on generality, all other values of the internal states 
$\{\sigma_1...\sigma_N\}$ can be obtained from that solution via renaming the
internal states accordingly.
Note that $M_i=0$ is explicitly allowed. Thus one can obtain solutions with
lower number of internal states from general ones with higher number of internal states.
For illustration, the example of the reduction of spin-$\frac12$ solutions
to spinless solutions is shown below.
Thus to obtain the energy eigenvalues of these models a
set of $N+M_1+M_2+\ldots+M_{N'}$ coupled equations for $N+M_1+M_2+\ldots+\!M_{N'}$
variables has to be solved. 
As already remarked, only the first $N$ variables of the first level of the solution are needed
to determine the energy eigenvalue.

For the rest of the paper we restrict ourselves to the case $N'=2$, 
namely two internal states, which can be interpreted as spin-$\uparrow$ and spin-$\downarrow$.
The explicit Bethe equations for this case reduce to four systems,
hardwall bosons, periodic bosons, hardwall fermions and periodic fermions,
which are (from top to bottom)
\be
\fl \qquad
\mathrm{e}^{\mathrm{i}2 k_i L}&=&
\prod_{j=1}^M
\frac{k_i  - \Lambda_j  - \frac12 \mathrm{i} c }
{k_i  -  \Lambda_j + \frac12 \mathrm{i} c}
\overbrace{\frac{k_i  + \Lambda_j -\frac12 \mathrm{i} c }
{k_i + \Lambda_j +\frac12 \mathrm{i} c}}^A
\prod_{j\neq i}^N
\frac{k_i - k_j + \mathrm{i}c}{k_i - k_j - \mathrm{i}c}
\overbrace{\frac{k_i + k_j + \mathrm{i}c}{k_i+k_j \!-\! \mathrm{i}c}}^A
\nonumber\\
\fl \qquad
\mathrm{e}^{\mathrm{i} k_i L}&=&
\prod_{j=1}^M
\frac{k_i  - \Lambda_j - \frac12 \mathrm{i} c}
{k_i - \Lambda_j + \frac12 \mathrm{i} c}
\prod_{j \neq i}^N
\frac{k_i - k_j + \mathrm{i}c}{k_i - k_j - \mathrm{i}c}
\nonumber\\
\fl \qquad
\mathrm{e}^{\mathrm{i}2 k_i L} &=&
\prod_{j=1}^M
\frac{k_i - \Lambda_j + \frac12 \mathrm{i} c }
{k_i - \Lambda_j - \frac12 \mathrm{i} c}
\,
\frac{k_i  + \Lambda_j + \frac12 \mathrm{i} c}
{k_i + \Lambda_j - \frac12 \mathrm{i} c}
\nonumber\\
\fl \qquad
\mathrm{e}^{\mathrm{i}k_i L}&=&
\prod_{j=1}^M
\frac{k_i - \Lambda_j + \frac12 \mathrm{i} c }
{k_i - \Lambda_j  - \frac12 \mathrm{i} c}.
\label{eq:BAE}
\ee
This set ($i=1,\ldots,N$) of $N$ 1st level Bethe equations
 is connected to a second set of so called spin roots $\{\Lambda_1,\ldots,\Lambda_M\}$ 
 obeying the $M$ second level Bethe equations
\be
&&
\prod_{j=1}^N
\frac{\Lambda_i - k_j	- \frac12 \mathrm{i} c }{\Lambda_i - k_j + \frac12 \mathrm{i} c	}
\overbrace{
\frac{\Lambda_i + k_j  - \frac12 \mathrm{i} c	}{\Lambda_i + k_j  + \frac12 \mathrm{i} c }}^A
=
\prod_{j=1 \atop j \neq i}^M
 \frac{\Lambda_i - \Lambda_j - \mathrm{i}c	}{\Lambda_i - \Lambda_j + \mathrm{i}c	}
\overbrace{ \frac{\Lambda_i + \Lambda_j - \mathrm{i}c	}{\Lambda_i + \Lambda_j + \mathrm{i}c	}}^{A}
\nonumber
\\
&&
\prod_{j=1}^N 
 \frac{
 \Lambda_i - k_j  - \frac12 \mathrm{i} c
}{
\Lambda_i - k_j  + \frac12 \mathrm{i} c
}
=
\prod_{j=1 \atop j\neq i}^M  
\frac{
 \Lambda_i - \Lambda_j  - \mathrm{i}{c}
}{
\Lambda_i - \Lambda_j  + \mathrm{i}{c}
}.
\label{eq:BAE-2ndlevel}
\ee

The 1st equation is for hard wall boundary systems, the 2nd equation is for the periodic models.
These 2nd level Bethe equations hold for both fermion and boson gases,
with the sign from $\Psi\to\pm\Psi$ under particle exchange entering only in the first level (\ref{eq:BAE}),
as can be seen from the details of the exact solution in \ref{appa}. 
Roots within one set have to be pairwise different and are usually complex numbers.
Here $M\leq{N}/{2}$ denotes the number of one of the spin components, e.g., $N_\uparrow=M$
and $N_\downarrow=N-M$. 
{}From the above definition it can be seen that the equations for a gas with $n$ internal degrees 
of freedom can be obtained from the solution with $n'>n$ internal degrees by taking states 
without values in these components. 
For example,  the above Bethe equations \eref{eq:BAE} and \eref{eq:BAE-2ndlevel} 
collapse for $M=0$ (blank all terms containing $\Lambda$'s) to
\be
\fl \qquad
&
\mathrm{e}^{\mathrm{i}2 k_i L} =
\prod_{j\neq i}^N
\frac{k_i - k_j + \mathrm{i}c}{k_i - k_j  -  \mathrm{i}c}
\,
{\frac{k_i + k_j + \mathrm{i}c}{k_i + k_j - \mathrm{i}c}}
\qquad  
\qquad  
&
\mathrm{e}^{\mathrm{i}2 k_i L} = 1  
\nonumber\\
\fl \qquad
&
\mathrm{e}^{\mathrm{i} k_i L} =
\prod_{j\neq i}^N
\frac{k_i - k_j + \mathrm{i}c}{k_i - k_j - \mathrm{i}c}
\qquad  
\qquad 
&
\mathrm{e}^{\mathrm{i} k_i L}\  = 1.
\ee
These are, respectively, the well known results for spinless bosons \cite{Gaudin} (left) and 
free fermions(right) with hard wall boundaries (1st line) and
spinless bosons \cite{LL} and free fermions with periodic boundary conditions (2nd line).
In each equation $i=1,\ldots,N$.

These equations are important, because the ground state of a Bose gas
is fully polarized \cite{periodicsu2bosons}, i.e., described by these spinless results \cite{LL,Gaudin}.
The terms denoted by A in equations \eref{eq:BAE} and \eref{eq:BAE-2ndlevel} are the new terms for the 
hard wall case.
They lead to the invariance of the Bethe equations under the exchange of any root 
$k_i \to -k_i$, $\Lambda^{(k)}_i\to - \Lambda^{(k)}_i$ for all states of the system.
From a physical point of view this means that the total momentum of the system is no longer conserved for hard wall boundaries, because reflection at the wall can change the direction 
but not the energy of a particle.

\section{Weak coupling expansions}
\label{sec:weak}
\begin{figure}[t]
	\begin{center}
 	 	\includegraphics[scale=0.3,angle=0]{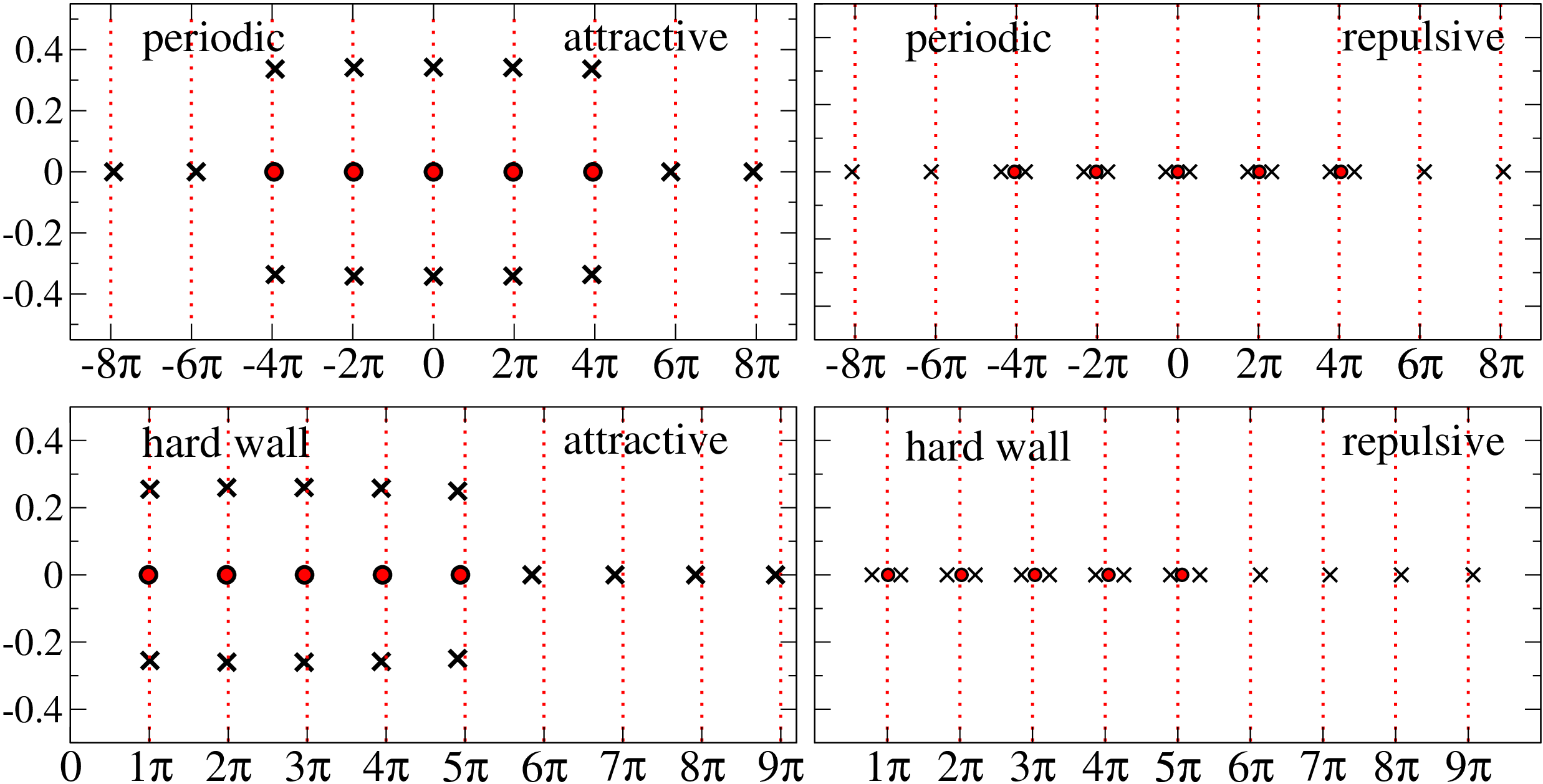}
\caption{
Generic root distribution  in the complex plane for the ground state for
weak attractive ($c=-1$, left) and weak repulsive coupling ($c=1$,right). 
Here $N=14, M=5$ and $L=1$, corresponding to five pairs of fermions
and four unpaired fermions, with $N_\downarrow = 9$ and $N_\uparrow=5$.
There is a continuous evolution from free particles at $c=0$, which are at 
multiples of $\pm\frac{2\pi}{L}$ for the periodic case  and $\frac{\pi}{L}$ for hard walls.
Circles denote $\Lambda_i$ and crosses denote $k_i$.
For each case the roots are obtained by solving the Bethe equations \eref{eq:BAE} and 
\eref{eq:BAE-2ndlevel}.
}
\label{fig:roots-weak-fermions}
	\end{center}
\end{figure}

In this section we discuss the spin-$\frac12$ fermionic system in the weak attractive and weak 
repulsive coupling limits via expansion of the Bethe equations.
For simplicity we concentrate on the ground state for a given polarization $M$, 
but many results and ideas hold as well for excitations.
An explicit analytic expression for the ground state energy is calculated from the linearized 
Bethe equations.

{}From numerical analysis of the Bethe equations it can be seen that the root distribution evolves 
continuously from the free fermions/bosons case $c=0$.
This implies that the momenta are still very close to the free momenta 
$k_i=\frac{2\pi}{L} n_i ,n_i=0,\pm 1,\pm 2,\ldots$ (periodic)
and $k_i=\frac{\pi}{L}n_i,n_i=1,2,3,\ldots$ (hard wall).
Note that $k_i=0$ is not a free ($c=0$) solution in the hard wall case, 
due to lack of translational invariance of the Hamiltonian, which has consequences
for setting up integral equations for the ground state, cf. 
equations \eref{eq:HW-int-eq-repulsive} and \eref{eq:HW-int-eq-attractive}.
For spin-$\frac12$ fermions at most two fermions can share one of these `lattice' sites, thus it 
is natural to interpret them as a pair of spin-$\uparrow$ and spin-$\downarrow$. 
Each lattice site can be either unoccupied, occupied by one $k_j$
or occupied by two quasi-momenta, in which case they are interspaced by a spin root 
$\Lambda$ (for the ground state only). 
In the attractive case these quasi-momenta form a complex pair $k_j=\alpha\pm \mathrm{i} \beta$ 
with real part $\alpha$ almost the same as the associated real $\Lambda$. 
A typical ground state root distribution is shown in \Fref{fig:roots-weak-fermions}.
The energy  $k_1^2+k_2^2=\alpha^2-\beta^2$ of such a pair is lower than the energy of the same 
two fermions in the limit $c=0$, suggesting the notion of `bound pair'. 
In the repulsive case the quasi-momenta are still interspaced by $\Lambda$, 
but lie on the real axis. 
Although there is no binding energy gained here, we will still use the term `bound pair' in the 
following to denote a pair of quasi-momenta $k_j$ associated with
the same site $\frac{2\pi}{L} n_i$ or $\frac{\pi}{L} n_i$.

The expansion of the Bethe equations follows similar work on systems containing  
bosons, fermions or mixtures of both in Batchelor \etal \cite{ours-hw-bosons,ours-TG,ours-fermions}.
The free parameter in the Bethe equations is $c L$, but for physical clarity we will expand in small $c$, 
thinking of $L$ as a fixed and finite number.
Before expanding the product and the exponential it is necessary to ensure that $x \gg c$ in $f(x)=\frac{x+\mathrm{i}c}{x-\mathrm{i}c}$ by separately treating
terms involving quasi-momenta and spin roots from the same lattice site $\frac{2\pi}{L}n_i$. 
Numerical analysis  readily establishes the Bethe roots to be of the form
\be
k^{(u)}_i=\frac{2\pi}{L} m_i + \delta_i^{(u)} c, \qquad 
\{m_i\}=\left\{\pm\frac{N-M-1}{2},\ldots,\pm\frac{M+1}{2} \right\}
\nonumber
\ee
for $i=1,\ldots,N-2M$
\be 
k^{(p)}_i=\frac{2\pi}{L} n_i + \delta_i^{(p)} c \pm \beta_i \sqrt{c},
\qquad
\{n_i\}=\left\{-\frac{M-1}{2},\ldots,\frac{M-1}{2}\right\}
\nonumber
\ee
for $i=1,\ldots,M$ and
\be
\Lambda^{(u)}_i= \frac{2\pi}{L}  n_i + \gamma_i c ,
\quad
\{n_i\}=\left\{-\frac{M-1}{2},\ldots,\frac{M-1}{2}\right\}.
\nonumber
\ee
Here superscripts $(p)$ and $(u)$ denote paired and unpaired roots 
with the real constants $\beta_i,\gamma_i,\delta_i$ to be determined.
For simplicity we consider only the case $N$ even and $M$ odd, thus
$m_i$ and $n_i$ always denote integers.
In the above the ground state configuration
for a given polarization $M$ has
all paired momenta lying symmetrically distributed closest to the origin with symmetrically 
packed unpaired momenta on an outside layer (recall \Fref{fig:roots-weak-fermions}).

Assuming this functional dependency of the Bethe roots to lowest order in $c$ it is possible 
to expand the Bethe equations consistently in powers of $\sqrt{c}$,  
leading to a set of linearly expanded equations to first order in $c$.
For the periodic case this leads to explicit expressions 
\be
\fl \qquad
k^{(u)}_i &=& \frac{2\pi}{L} m_i  +  {c} \sum_{j=1}^M \frac{1}{2\pi(m_i - n_j)}
\nonumber
\\
\fl \qquad
k^{(p)}_i &=&
\frac{2\pi}{L} n_i \,
\pm \sqrt{\frac{c}{L}}
+
{c}
\left[ 
{
\sum_{j\neq i }^M \frac{1}{2\pi(n_i - n_j)}
 +
\frac12  \sum_{j=1}^{N-2M} 
\frac{1}{2\pi(n_i - m_j)}
}
\right]
\ee
for the location of the Bethe roots.
Similarly, for the hard wall case 
\be
\fl
&&k^{(u)}_i =
\frac{\pi}{L} m_i
+
\frac{c}{2\pi } \sum_{j=1}^M \left(
 \frac{1}{m_i - n_j}
+
 \frac{1}{m_i + n_j}
\right)
\nonumber\\
\fl
&&k^{(p)}_i =
\frac{\pi}{L} n_i \,
\pm \sqrt{\frac{c}{2L}}
+
\frac{c}{2\pi}
\left[ 
\sum_{j\neq i }^M 
\left(
\frac{1}{n_i - n_j}
+
\frac{1}{n_i + n_j}
\right)
\right.
\nonumber\\
&&\qquad\qquad \left. 
+
 \frac{1}{2}
\sum_{j=1}^{N-2M} 
\left(
\frac{1}{n_i - m_j}
+
\frac{1}{n_i + m_j}
\right)
+
\frac{1}{ n_i}
\right]
\ee
Note the $1/n_i$ term, which is a feature of the hard wall case.
This correction has a significant
effect and is also found in the equivalent Hubbard model expansion
\cite{Michael-Hubbard-preprint}.
It contributes to the difference 
in energy for periodic and hard wall models in the thermodynamic limit, the so called surface energy.

\begin{figure}[t]
	\begin{center}
  	 	\includegraphics[scale=0.32,angle=0]{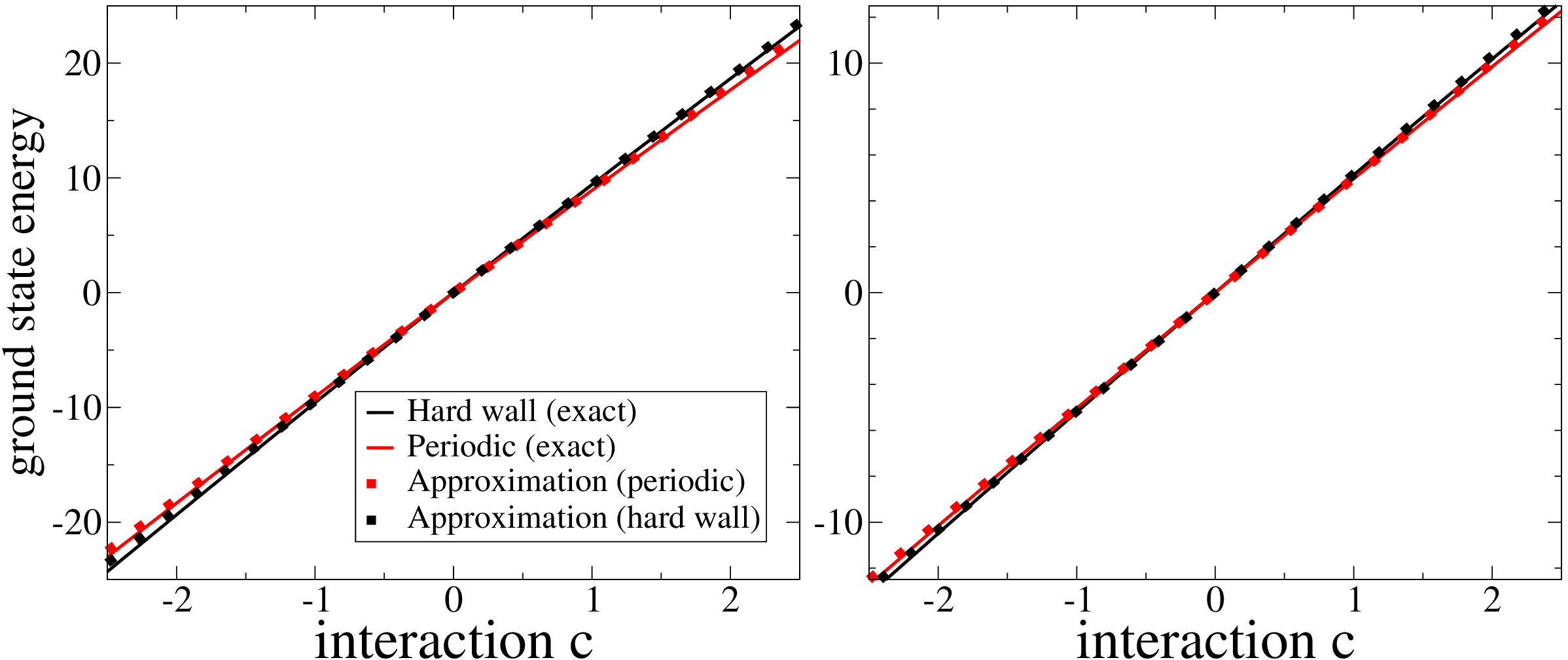} 
\caption{
Comparison for the ground state energy per particle between the analytic weak coupling results 
\eref{eq:weak-gs-analytic-PBC} and \eref{eq:weak-gs-analytic-HW} and numerical
values obtained by solving the full Bethe equations.
Here $N=18$ with $M=9$ (left) and $M=3$ (right).
For each case the energy is given without the free particle contribution, for ease of comparison. 
}
 		\label{fig:weak-gs-comparison}
	\end{center}
\end{figure}

In both cases the binding strength, especially in the attractive regime, is the same for all pairs, 
which is found to be true numerically only for weak coupling~\cite{Gaudin}.
For stronger coupling the inner lying pairs appear to be more strongly bound 
(i.e., they have a larger imaginary part ) than the outer lying pairs, 
which is not visible in low order of $c$.
It eventually leads to a break down of the weak coupling expansion for stronger interaction.
Summing over the squares of the approximated roots $k_i$, 
the ground state energy to first order in $c$ for the periodic case
is given by \cite{ours-fermions}
\be
\fl \qquad
E_{\rm PBC}
= \left(
\frac{2\pi}{L}\right)^2
\left[
\case{1}{12} N^3
-\case14{M} N^2
+
\left(
\case14{M^2}-\case{1}{12}\right) N
\right]
+
\frac{2c}{L}
(N-M) M.
\label{eq:weak-gs-analytic-PBC}
\ee
Similarly for the hard wall case
\be
\fl\qquad
E_{\rm HW}
= 
 \left(
\frac{\pi}{L}
 \right)^{2}
\left[
\case13 N^3 
+
\left(\case12 - M \right) N^2
+
\left(M^2 - M + \case{1}{6}\right)
N
+
M^2 
\right]
\nonumber\\
\qquad\qquad
+
\frac{c}{L}
2 (N-M+1) M.
\label{eq:weak-gs-analytic-HW}
\ee
\Fref{fig:weak-gs-comparison} shows a comparative plot for the ground state energy 
for two different fillings in the weak attractive and weak repulsive regimes.
 The exact result
(obtained numerically) is compared with the analytical approximations 
\eref{eq:weak-gs-analytic-PBC} and \eref{eq:weak-gs-analytic-HW}.

\section{Strong coupling expansions}
\label{sec:hard}

\begin{figure}
	\begin{center}
	 	\includegraphics[scale=0.34,angle=0]{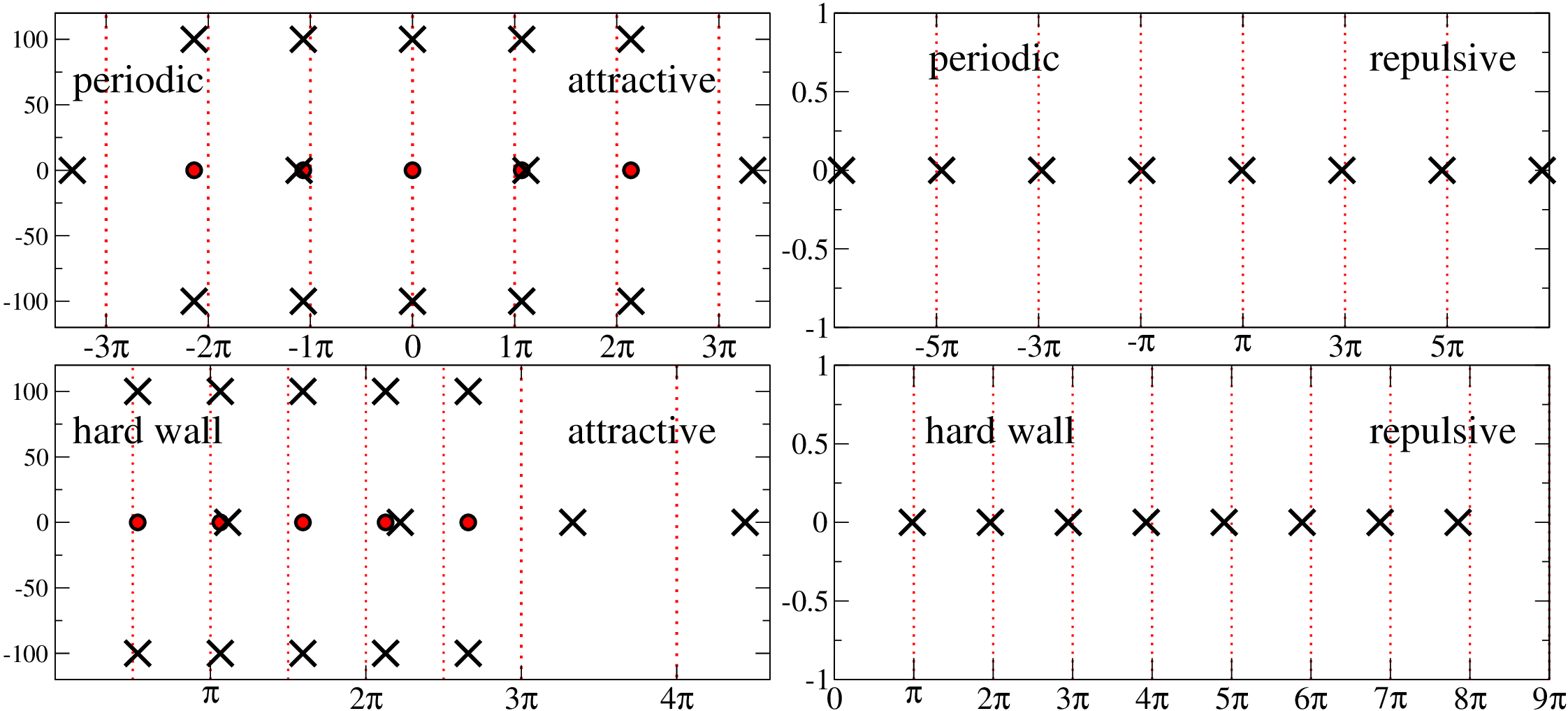} 
\caption{
The ground state quasi-momenta distribution for periodic and hard wall fermions in the strong coupling
regime $c\to \pm \infty$ for different polarization $M$.
{(top left) }{Strongly attractive periodic case.} 
Complex conjugate pairs of quasi-momenta $k$ are approximately interspaced by spin roots $\Lambda$ with $k=\Lambda\pm\mathrm{i}\frac12{c}$.
{(bottom left) }{Strongly attractive hard wall case.}
Similar to the attractive periodic case.
{(top right) }{Strongly repulsive periodic case.} 
Here the spin roots $\Lambda$ wander out proportionally with c towards $\pm\infty$ 
but remain real \cite{Takahashi}. 
The quasi-momenta $k_i$ break up the `bound pairs' and tend to
their tightly bound fermion values $k_i=\frac{\pi}{L} n_i$,$n_i=\pm1,\pm3,\ldots$.
{(bottom right) }{Strongly repulsive hard wall case.}
Similar to the repulsive periodic case. 
The spin roots $\Lambda$ wander to $+\infty$,
but the quasi-momenta tend to the tightly bound fermion values 
$k_i=\frac{\pi}{L}n_i$,$n_i=1,2,\ldots$ breaking up `bound pairs' (see text).
In each figure the quasi-momenta $k$ and the spin roots $\Lambda$ are denoted by crosses 
and circles where applicable.
The numerical values are $N=8, M=3, c=500, L=1$ (repulsive cases) and 
$N=14, M=5, c=-200, L=1$ (attractive cases).
The dotted lines are guides to the eye.
}
 		\label{fig:roots-fermions}
	\end{center}
\end{figure}

\subsection{Strong repulsive limit $c\to\infty$}

Consider first the periodic case.
As the repulsive strength of interaction is increased the quasi-momenta defining the ground state
move away from the free values $k_i=\frac{2\pi}{L} n_i, n_i=0,\pm 1,\pm 2, \ldots$ 
towards the values $k_i=\frac{\pi}{L} n_i, n_i=\pm1,\pm 3,\ldots$ 
for tightly bound fermions.
For the hard wall case the quasi-momenta return to the free values  
$k_i=\frac{\pi}{L} n_i, n_i=1,2,\ldots$. 
The spin roots $\Lambda$ wander out to infinity linearly with $c$, except for at most
one $\Lambda=0$ in the periodic case \cite{Takahashi}. 
For the hard wall case the behavior is similar. 
An illustration of the root behavior for the different cases is given in \Fref{fig:roots-fermions}. 
For the ground state all roots remain real.
The energy is given in terms of a set of $M$ parameters $\{\gamma_i\}$, 
appearing as the leading coefficients in the spin roots $\Lambda_i$.

In this limit, the roots are given in integer powers of $c$, in agreement with numerical checks. 
For the periodic repulsive case the ground state roots are given by
(see also chapter 2.27 of Ref. \cite{Takahashi}) 
\be
\fl \qquad
k_i=\frac{\pi}{L} n_i 		+ {\delta_i} {c}^{-1}  
\qquad
&
 i=1,\ldots,N
&
\qquad
\{n_i\}=\{
\pm 1,\pm 3,\ldots,\pm (N-1)   \}
\nonumber\\
\fl \qquad
\Lambda_i =   \gamma_i c  
\qquad
&
i=1,\ldots,M
&
\qquad
 \textrm{(with one $\Lambda=0$ for $M$ odd)}
\nonumber
\ee
where we show only orders which are relevant for the energy to order $c^{-1}$.
The real constants $\delta_i$ and $\Lambda_i$ are to be determined.
Expanding the second level Bethe equations up to order $c^{-1}$
the parameters $\gamma_i$ are found to satisfy the same parameter-free equations 
\be
\left(
\frac{\gamma_i +\frac12\mathrm{i} }{\gamma_i -\frac12\mathrm{i}}
\right)^N=-\prod_{j=1}^M \frac{\gamma_i -\gamma_j + \mathrm{i}  }{\gamma_i -\gamma_j + \mathrm{i} }
 \quad
 \qquad
 \textrm{for  }
i=1,\ldots,M
\label{eq:HB-chain-eq}
\ee
as for the isotropic Heisenberg spin chain.
From the expansion of the first level Bethe equation up to order $c^{-1}$ 
the roots are found to be (for $i=1,\ldots,M$)
\be
k_i
= \frac{\pi}{L} n_i \left(1- \frac{1}{cL} \Omega^N_M \right).
\ee
Note that the constant $\Omega^N_M=\sum_{j=1}^M \frac{1}{\gamma_i^2 + 1/4}$ depends only
on the set $\{\gamma_i \}_{i=1,\ldots,M}$ and neither on the interaction $c$ nor on the 
integers $n_i$, making numerical evaluation 
significantly easier than for the initial Bethe equations.

\begin{figure}[t]
	\begin{center}
	 	\includegraphics[scale=0.25,angle=0]{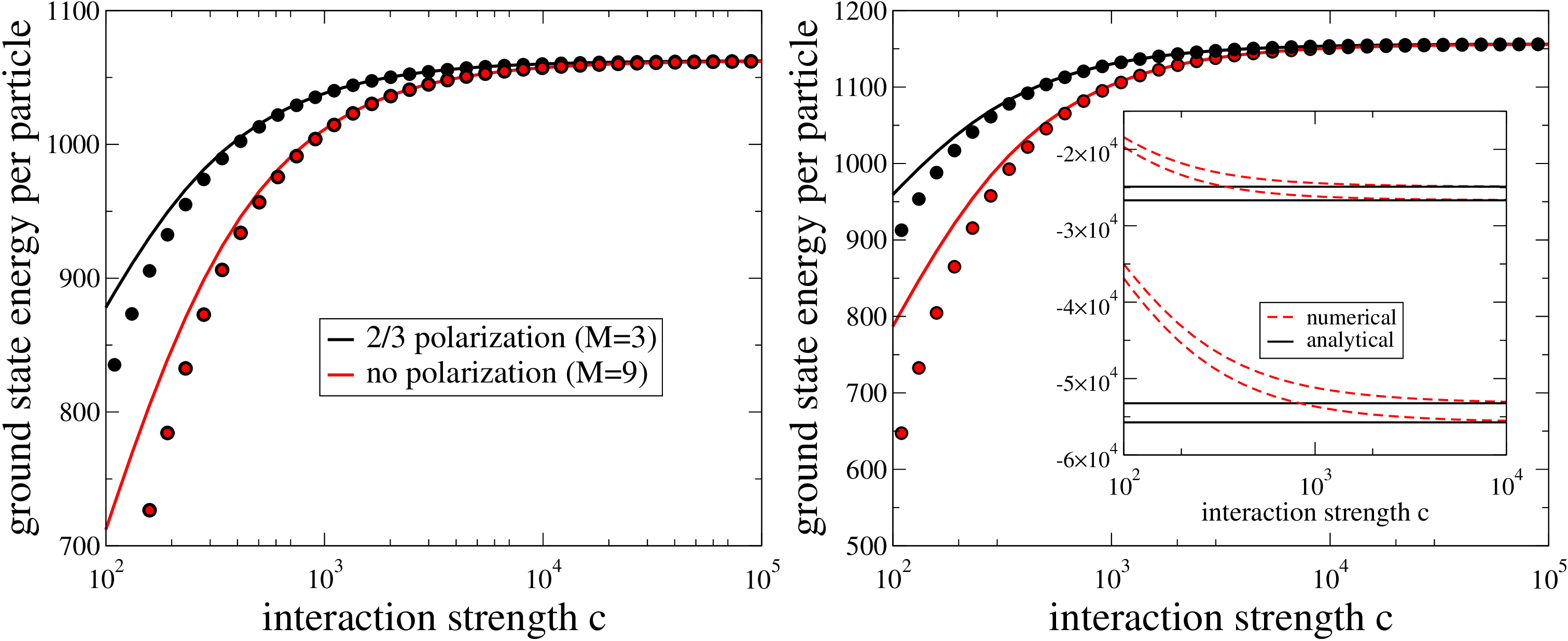} 
\caption{
Comparison of	the ground state energy in the strong repulsive 
interaction limit $c\to \infty$ for (left) periodic boundaries and (right) hard walls.
The circles denote the values obtained from the leading terms evaluated in 
equations \eref{eq:strong-PBC-rep-GS} and \eref{eq:strong-HW-rep-GS}.
The full curves are the results obtained from the exact numerical solution of the 
Bethe equations.
The inset shows confirmation of the order $c^{-1}$ coefficient.
Dashed lines denote numerical values obtained from 
$(E_\textrm{\tiny num}-E_\textrm{\tiny asympt})\cdot c$ while full lines are
the $c^{-1}$ coefficients from equations \eref{eq:strong-PBC-rep-GS} and \eref{eq:strong-HW-rep-GS}.
The lines converge to the same value for large $c$.
The four cases shown are as in the main panel.
The same numerical values are used as in \Fref{fig:weak-gs-comparison}, 
namely $N=18$ and $M=3,9$ with $L=1$.
}
 		\label{fig:strong-repulsive-approx-gs}
	\end{center}
\end{figure}

The energy expression up to order $c^{-1}$ follows as
\be
E_{\rm PBC}=
\frac{\pi^2  }{3 L^2}
(N^3-N)
 \left[1- \frac{2}{c L} \Omega_M^N 
+
O(c^{-2})
\right]
\label{eq:strong-PBC-rep-GS}
\ee
where we have taken the integers $n_i$ for the ground state configuration.
Comparisons between this expansion and the exact numerical solution 
for finite systems are given in \Fref{fig:strong-repulsive-approx-gs}.
Formulas for  the hard wall case are derived in a similar fashion.
Here
\be
k_i=\frac{\pi}{L} n_i + \delta_i c^{-1} ,
&
i= 1,\ldots,N
&
\qquad
\{n_i\} = \{  1,2,\ldots,N \}
\nonumber\\
\Lambda_i=  \gamma_i\  c^1, 
\qquad\qquad\qquad
&
  i = 1,\ldots,M
&
\nonumber
\ee
in the strong coupling limit $c\to\infty$.
Expansion of the second level Bethe equations results in the spin root 
coefficients $\gamma_i$ satisfying 
\be
\left(
\frac{\gamma_i+ \frac12 \mathrm{i}  }{\gamma_i- \frac12\mathrm{i}}
\right)^{2N}
=
\prod_{j\neq i}^M
\frac{\gamma_i - \gamma_j + \mathrm{i} }{\gamma_i - \gamma_j - \mathrm{i}}
\,
\frac{\gamma_i + \gamma_j + \mathrm{i} }{\gamma_i + \gamma_j - \mathrm{i}}
\label{eq:openXXX}
\ee
for $i=1,\ldots,M$.
These are the well known Bethe equations for the open Heisenberg chain \cite{Gaudin,ABBBQ}.
Expansion of the first level Bethe equations to order $c^{-1}$ gives the quasi-momenta $\{k_i\}$ 
depending only on the parameters $\{\gamma_i\}$ via
\be
k_i 
=\frac{\pi}{L} n_i 
\left(
1- \frac{1}{c L}
\hat{\Omega}_M^N
\right).
\ee
Here the constant $\hat{\Omega}_M^N=\sum_{j=1}^M \frac{1}{\gamma_i^2+1/4}$ 
is obtained from the solution $\{\gamma_i\}_{i=1,\ldots,M}$ of equations \eref{eq:openXXX}.
The leading terms in the ground state energy follow as
\be
E_{\rm HW}=
\frac{\pi^2  }{6 L^2}
\left(
2 N^3 + 3 N^2 + N
\right)
\left[1- \frac{1}{c}\frac{2}{L} \hat{\Omega}_M^N 
+
O(c^{-2})
\right].
\label{eq:strong-HW-rep-GS}
\ee
A comparison for the ground state energy between this expansion and
the exact numerical result is shown in \Fref{fig:strong-repulsive-approx-gs} 
for a finite system with different polarizations.
The results are in excellent agreement with the numerical curve as soon as the assumption
$\Lambda\sim c$ holds, which happens for certain system sizes $N$ and $M$ at large 
interaction strength $c>1000$. 
Note that here we have plotted against the interaction strength $c$, one can also do similar 
plots against the rescaled interaction $\gamma$.

\begin{figure}[t]
	\begin{center}
	 	\includegraphics[scale=0.3,angle=0]{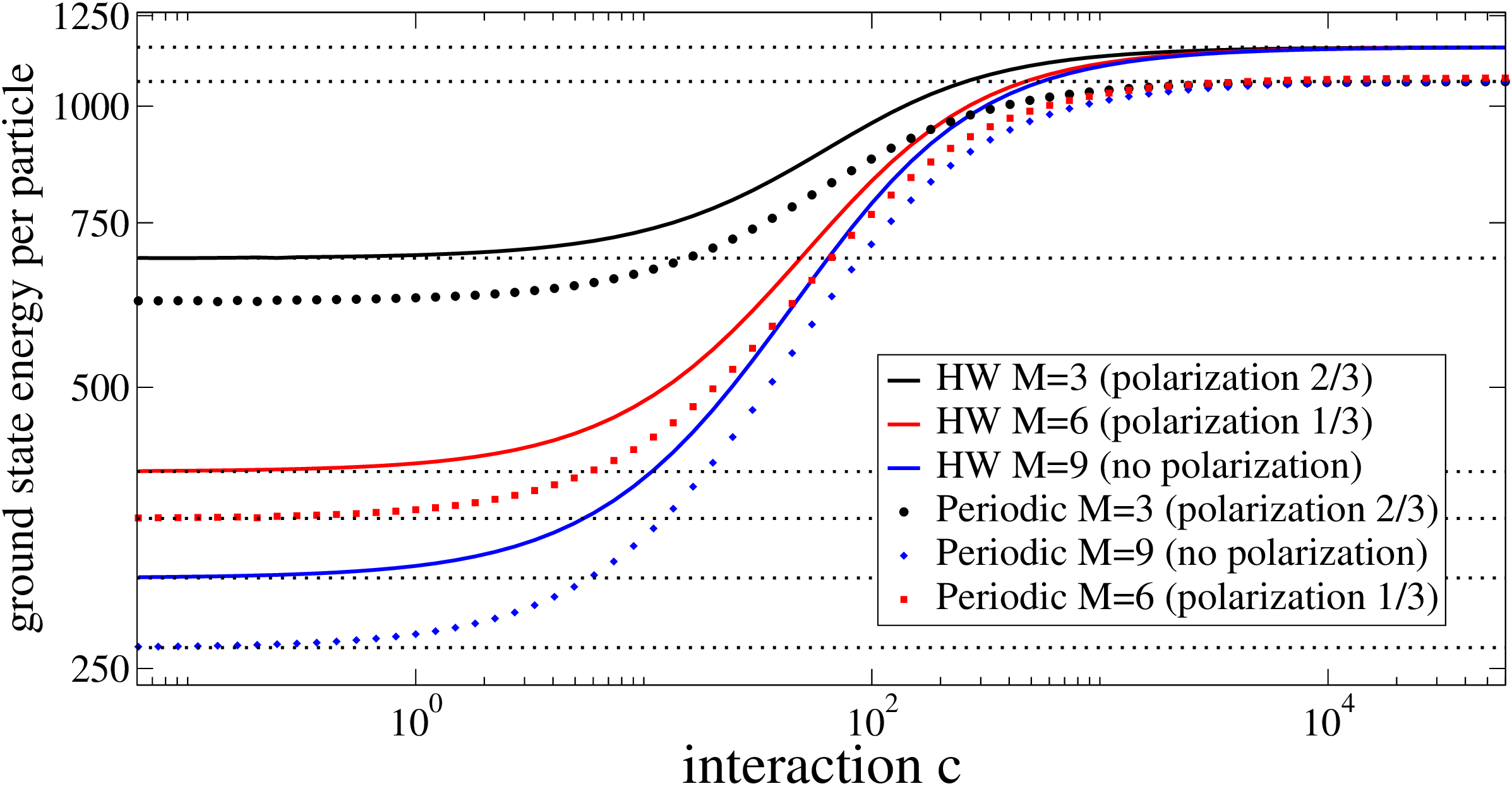} 
		\caption{
Ground state energy per particle for repulsive periodic and hard wall fermions 
versus interaction strength $c$.
The dotted lines denote the values for free $c=0$ and tight binding $c\to\infty$ fermions. 
Full polarization corresponds to non-interacting fermions and is thus only a horizontal line (not shown).
For given polarization the periodic system always has a lower energy than the corresponding
hard wall system.
Shown are numerical values for $N=18$ and $M=3,6,9$ with $L=1$.
}
 		\label{fig:Fig1}
	\end{center}
\end{figure}
%
%
\Fref{fig:Fig1} shows a double logarithmic plot of
the ground state for a generic finite system 
with different polarization, obtained numerically from the original Bethe equations.
The dotted lines denote the known asymptotic values for free ($c\to 0$) and tightly bound
fermions ($c\to\infty$), which agree with the leading order of the expansions derived here.
Note that for a finite size system the hardwall case has a higher energy than the periodic case.
\subsection{Strong attractive limit $c\to -\infty$}

In this section we again for convenience consider the case $N$ even with $M$ odd, 
the generalization from which should be clear.
The root configuration for the ground state consists of $2M$ complex conjugated 
quasi-momenta $k$, each pair interspaced by a real spin root $\Lambda$. 
The remaining $N-2M$ unpaired momenta $k$ lie on the real axis.
The occurrence of complex roots makes the analysis more complicated. 
Following Takahashi we assume $k_{i}=\Lambda_i \pm \frac12 \mathrm{i}{c}$
for the $2M$ paired quasi-momenta. 
This is a good approximation for the thermodynamic limit \cite{Takahashi}.
We assume here that it is also a good approximation
for large attractive interaction strength $c\to -\infty$. 
Using this form the Bethe equations can be rewritten (see chapter 2.2.8 of Ref.~\cite{Takahashi})  
in terms of the root sets
$\{k_i\}_{i=2M+1,\ldots,N}$ and $\{\Lambda_i\}_{i=1,\ldots,M}$, i.e., consisting only of the remaining
(real) quasi-momenta and the (real) spin roots $\Lambda_i$, as
\be
\fl \quad 
{\mathrm e}^{\mathrm{i} k_i L}=\prod_{j=1}^M 
\frac{k_i  -  \Lambda_j  +  \frac12 \mathrm{i}{c}}{k_i  - \Lambda_j  - \frac12 \mathrm{i}{c}},
 \quad
{\mathrm e}^{\mathrm{i}2 \Lambda_i L}
 = 
     \prod_{j=2M + 1}^N 
\frac{\Lambda_i  -  k_j  +  \frac12 \mathrm{i}{c} }{\Lambda_i  -  k_j  -  \frac12 \mathrm{i}{c}}
\prod_{j\neq i}^M 
\frac{\Lambda_i  -  \Lambda_j  +  \mathrm{i}c  }{\Lambda_i  -  \Lambda_j  -  \mathrm{i}c}.
\label{eq:modified-BAE-PBC} 
\ee
All results in this subsection refer to the analysis and numerical evaluation of these 
modified Bethe equations.
Asymptotic and numerical analysis suggests that the roots have the form ($n_i,m_i\in \mathbb{Z}$)
\be
\fl \qquad
k_i = m_i \frac{\pi}{L}   + \frac{\delta_i}{c},  
&\quad i=2M+1,\ldots,N, &\quad 
m_i=
\{ 
\pm 1,\pm 3,\ldots,\pm (N-2M-1) 
\} 
\nonumber\\
\fl \qquad
\Lambda_i = n_i \frac{\pi}{L}   + \frac{\gamma_i}{c},  
&\quad i=1,\ldots,N, &\quad n_i=\left\{ -\frac{M-1}{2},\ldots,\frac{M-1}{2} \right\}. 
\nonumber
\ee
Expansion in the small parameter $c^{-1}$ results in
\be
k_i &=& 
m_i \frac{\pi}{L}  
\left( 1 - \frac{4M}{L c} \right)
\nonumber \\
\Lambda_i &=&
n_i \frac{\pi}{L}
\left(
1-\frac{2N-3M}{L c} 
\right).
\ee
This root distribution gives the energy 
\be
\fl \quad
E_{\rm PBC}&=& 
\sum_{i=1}^M \left( 2\Lambda_i^2 -\frac{c^2}{2}  \right) + \sum_{i=2M+1}^N k_i^2
\nonumber\\
\fl \quad
&=&
-\case12{M}c^2
+
\frac{\pi^2}{L^2}
\left\{
2\left(
1-\frac{4N - 6M}{Lc}
\right)
\frac{M^3-M}{12}
\right.
\nonumber\\
\fl \quad\qquad
&& 
\left.
+\left(
1-\frac{8M}{Lc}
\right)
\case13 \left[
{N^3 -6 M N^2 
+ 
(12 M^2 -1) N + 2 M - 8M^3}\right]
\right\}
\nonumber\\
\fl\quad\qquad
&& 
+ O(c^{-2}). 
\label{eq:strong-attractive-gs-approx-PBC}
\ee

The derivation for the hard wall case follows the same steps, with
the corresponding modified Bethe equations
\be
\fl \quad
{\mathrm e}^{\mathrm{i} 2 k_i L}=\prod_{j=1}^M 
\frac{k_i  -  \Lambda_j  +  \frac12 \mathrm{i}{c} }{k_i  - \Lambda_j  -  \frac12 \mathrm{i}{c}}
\frac{k_i  +  \Lambda_j  + \frac12 \mathrm{i}{c}  }{k_i  + \Lambda_j  -  \frac12 \mathrm{i}{c}},
\qquad i=2M+1,\ldots,N
\nonumber \\
\fl \quad
{\mathrm e}^{\mathrm{i}4 \Lambda_i L}
 = 
     \prod_{j=2M + 1}^N 
\frac{\Lambda_i  -  k_j  +  \frac12 \mathrm{i}{c}  }{\Lambda_i  -  k_j  - \frac12 \mathrm{i}{c}}
\frac{\Lambda_i  +  k_j  +  \frac12 \mathrm{i}{c}  }{\Lambda_i  +  k_j  - \frac12 \mathrm{i}{c}}
\, \prod_{j\neq i}^M 
\frac{\Lambda_i  -  \Lambda_j  +  \mathrm{i}c  }{\Lambda_i  -  \Lambda_j  -  \mathrm{i}c}
\frac{\Lambda_i  +  \Lambda_j  +  \mathrm{i}c  }{\Lambda_i  +  \Lambda_j  -  \mathrm{i}c}
\label{eq:modified-BAE-HW}
\ee
for $i=1,\ldots,M$.
Expanding these equations to order $c^{-1}$ leads to
\be
\fl \quad
k_i= m_i \frac{\pi}{L}
\left(
1 - \frac{4}{L c} M
\right)
& \quad
i=2M+1,\ldots,N 
& \quad
m_i=\{ 1,2,\ldots,N-2M \}
\nonumber\\
\fl \quad
\Lambda_i= n_i \frac{\pi}{L}
\left(
1 - \frac{2N-3M-1}{L c} 
\right)
& \quad
i=1,\ldots,M 
&\quad
n_i=\left\{ \frac12,1,\frac32,\ldots,\frac{M}{2} \right\}
\nonumber
\ee
Using these expressions for the roots gives the expansion for the energy 
\be
\fl \quad
E_{\rm HW}&=&
-\case12{M}c^2
+
\frac{\pi^2}{L^2}
\left[
\case{1}{12}\left(
1-
\frac{4N -
 6M-2}{Lc}
\right)
{M (M+1)(2M+1)}
\right.
\nonumber\\
\fl
\quad
&& +
\left.
\left(
1-\frac{8M}{Lc}
\right)
\case16 (N-2M) (N-2M+1)(2N-4M+1)
\right]
+ O(c^{-2}).
\label{eq:strong-attractive-gs-approx-HW}
\ee

\begin{figure}[t]
	\begin{center}
	 	\includegraphics[scale=0.24,angle=0]{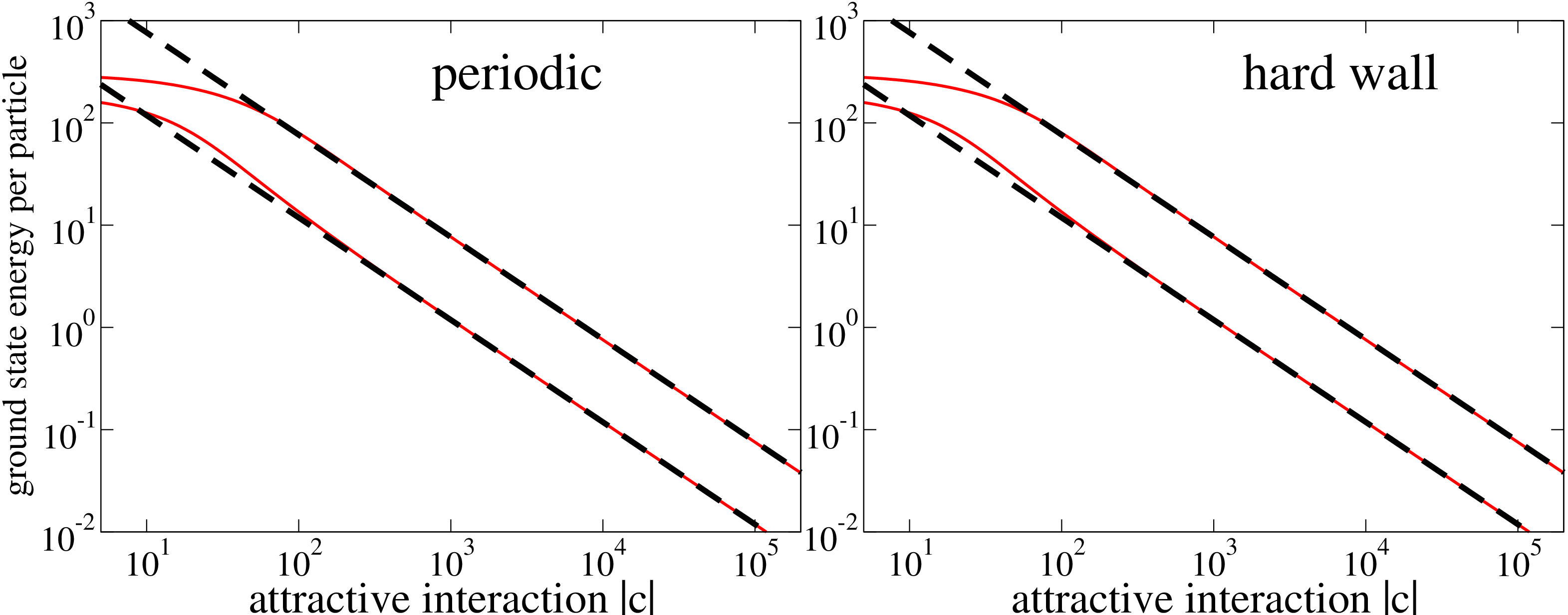} 
		\caption{
Ground state energy in the strong attractive region.
The curves show the comparison between the numerical solution of 
the modified Bethe equations \eref{eq:modified-BAE-PBC} and \eref{eq:modified-BAE-HW}[solid lines] 
and the expansions \eref{eq:strong-attractive-gs-approx-PBC} and \eref{eq:strong-attractive-gs-approx-HW}[dashed lines].
For clearer comparison both the binding energy $E_\textrm{bind}\sim c^2$ of the paired quasi-momenta
and the $O(c^0)$ energy term is subtracted.
The values used are $N=18$ with $M=3$ (upper curves) and $M=9$ (lower curves) with $L=1$. 			
}
 		\label{fig:comparo-strong-gs}
	\end{center}
\end{figure}

\Fref{fig:comparo-strong-gs} shows a comparison between the expansions 
\eref{eq:strong-attractive-gs-approx-PBC} and \eref{eq:strong-attractive-gs-approx-HW} 
and the exact ground state energy for the strong attractive case for periodic and hard wall
boundary conditions. 
Note that both the expansion and the exact numerical solution refer to the modified Bethe equations 
with paired quasi-momenta of the form $k=\Lambda \pm \frac12 \mathrm{i}{c}$.

\section{Integral equations for the ground state energy}
\label{sec:NLIE}

It is convenient to introduce the standard logarithmic form 
(see, e.g., Ref.~\cite{Takahashi}) of the Bethe equations \eref{eq:BAE} and 
\eref{eq:BAE-2ndlevel}.
In terms of the function $\phi(x)=\arctan (x/c)$, the equations for periodic fermions are
\be
k_i L = 2\pi I_i  - 
2 \sum_{j=1}^M \phi \left(2(k_i -\Lambda_j)\right)
\nonumber
\\
2 \sum_{j=1}^N \phi \left(2(\Lambda_i-k_j)\right) 
=
2\pi J_i  +
2 \sum_{j=1}^M \phi (\Lambda_i -\Lambda_j) 
\ee
and for hard wall fermions 
\be
\nonumber
2 k_i L = 2\pi I_i  - 2 \sum_{j=1}^M 
\left[ \phi ( 2(k_i -\Lambda_j))+
\phi \left(2 (k_i +\Lambda_j) \right)
\right]	
\\
%
2 \sum_{j=1}^N 
\left[
\phi  \left( 2(\Lambda_i - k_j) \right) 
+
\phi  \left( 2(\Lambda_i + k_j) \right) 
\right]
\nonumber \\
\qquad = 2\pi J_i 
+
2 \sum_{j\neq i}^M 
\left[
\phi  (\Lambda_i - \Lambda_j)  
+
\phi (\Lambda_i + \Lambda_j)  
\right].
%
%
\ee
The constants $I_1,\ldots,I_N$ and $J_1,\dots,J_M$ act as quantum numbers, 
they are either all  integer or half-odd integer within one set, depending
on the boundary conditions and if $N$ and $M$ are even or odd \cite{Takahashi}.
Generally the solution for a given allowed set of quantum numbers is unique
\cite{Korepin}.
The ground state for the fermionic system is given by the `half-filling' $M={N}/{2}$ \cite{Takahashi},
while in comparison the ground state of bosonic systems of this type is totally polarized ($M=0$) \cite{periodicsu2bosons}.
As a result, the ground state analysis of  bosonic systems 
can be done with the one-level Bethe equations for spinless bosons (given above), as long
as no special polarization is required.

We only consider here the special case of fermionic systems with $N$ even and $M$ odd.
The quantum numbers $I_n$ and $J_n$  for the ground state configuration are all one unit away 
from each other and either integers or half-odd integers, as they result from counting phases.
For excited states the quantum numbers become more spaced. 
For the ground state in the attractive case a more convenient, 
approximative form (see eqns 2.115 and 2.116 in Ref.~\cite{Takahashi}
as well as \eref{eq:modified-BAE-PBC},\eref{eq:modified-BAE-HW}
in the previous section)
of the Bethe equations in which all roots are real can be used. 
Analogous equations hold for the hard wall case.
From these equations the behavior in the thermodynamic limit can be obtained
via the standard integral equation procedure \cite{Gaud,Yang,Takahashi,Korepin,Gaudinbook}.
In the limit $N,L\to \infty$ with constant particle  density $N/L$
it is easiest to consider only the densities $\rho(k)$ and $\sigma(\Lambda)$ of the root sets
$\{ k_i \}$ and $\{\Lambda_i\}$, which are obtained from knowledge of the distribution
of the quantum numbers $I_n$ and $J_n$ in the various regimes and for the different
boundary conditions.

In the repulsive periodic case the ground state roots are all real and their density is 
determined by the coupled equations
\be
\rho(k) &=&
\frac{1}{2\pi}
+ \frac{2c}{\pi}\int_{-B}^B \frac{ \sigma(\Lambda) \textrm{d}\Lambda }{c^2 + 4 (k-\Lambda)^2}
\nonumber
\\
\sigma(\Lambda) 
&=&
-\frac{c}{\pi}\int_{-B}^B \frac{ \sigma(\Lambda') \textrm{d}\Lambda' }{c^2 +  (\Lambda-\Lambda')^2}
+
\frac{2c}{\pi}\int_{-Q}^Q \frac{ \rho(k) \textrm{d}k }{c^2 + 4 (k-\Lambda)^2}
\ee
The density $n$ and the polarization $n_\downarrow$ determine the constants $B$ and $Q$ 
in the equations below, making it possible to calculate the energy per particle $e$, with
\be
e=\int_{-Q}^Q k^2 \rho(k) \textrm{d}k, 
\quad
n=\int_{-Q}^Q  \rho(k) \textrm{d}k, 
\quad 
n_\downarrow=\int_{-B}^B  \sigma(\Lambda) \textrm{d}\Lambda. 
\ee

For the attractive case, the possibility of bound pairs changes the form of these equations to
\be
\rho(k) = 
&
\frac{1}{2\pi}
& 
+
\frac{2c}{\pi}\int_{-B}^B \frac{ \sigma(\Lambda) \textrm{d}\Lambda }{c^2 + 4 (k-\Lambda)^2}
\nonumber
 \\
\sigma(\Lambda) 
=
&
 \frac{1}{\pi}
+
&
\frac{c}{\pi}\int_{-B}^B \frac{ \sigma(\Lambda') \textrm{d}\Lambda' }{c^2 +  (\Lambda-\Lambda')^2}
+
\frac{2 c}{\pi}\int_{-Q}^Q \frac{ \rho(k) \textrm{d}k }{c^2 + 4 (k-\Lambda)^2}
\ee
with the auxiliary conditions
\be 
\nonumber
n= 2 \int_{-B}^B  \sigma(\Lambda) \textrm{d}\Lambda 
+\int_{-Q}^Q  \rho(k) \textrm{d}k, 
\qquad 
n_\downarrow-n_\uparrow=\int_{-Q}^Q  \rho(k) \textrm{d}k 
\\
e= 
\int_{-B}^B \left(2 \Lambda^2 - \frac12 {c^2}\right) \sigma(\Lambda) \textrm{d}\Lambda 
+
\int_{-Q}^Q k^2 \rho(k) \textrm{d}k.
\ee

These equations have been extensively examined in the periodic case 
(see, e.g., \cite{Wadati-expansions,ours-fermions} and references therein).
To obtain the integral equations for the hard wall case it is necessary to proceed 
with more caution --
the lack of translational invariance forbids the roots to take the value zero, 
even though it leads to a solution of the Bethe equations.
It is convenient to symmetrize the exponential Bethe equations to include sets of
zeros of the form $\{ \pm k_i,0\}$ and $\{\pm\Lambda_j ,0 \}$\ ~\cite{Gaudinbook,Fujimoto}.
This makes the equations somewhat resemble a special system of $2N+1$ particles 
(see also Ref.~\cite{Gaudinbook} for a remark on the equivalence of spinless bosons 
with hard walls and periodic boundary conditions).
The procedure then becomes analogous to the periodic case (e.g., see chapters 
2.2.6-2.2.8 in Ref.~\cite{Takahashi} and~\cite{ours-hw-bosons}).
It has, e.g., been applied to the
XXZ chain~\cite{BortzXXZ} and the
 open end Hubbard model \cite{KluemperHubbardbook,Michael-Hubbard-preprint}.
Note that the auxiliary integrals contain additional factors and terms accounting for the zero value.
The resulting integral equations for repulsive interactions and hard walls are
\be
\fl
\nonumber
\rho(k) 
&=& 
\frac{1}{\pi}
+ \frac{2c}{\pi}
\int_{-B}^{B} \frac{
\sigma(\Lambda)
\textrm{d}\Lambda
}{c^2+4(k-\Lambda)^2}
-
{
{
\ \frac{1}{L}
\frac{2c}{\pi}
\frac{1}{c^2 + 4 k^2}
\
}
}
\\
\fl
\sigma(\Lambda) 
&=& 
%
%
-\frac{c}{\pi}
\int_{-B}^{B} \frac{
\sigma(\Lambda')
\textrm{d}\Lambda'
}{c^2+(\Lambda-\Lambda')^2}
+
\frac{2c}{\pi}
\int_{-Q}^{Q} \frac{
\rho(k)
\textrm{d}k
}{c^2+4(k-\Lambda)^2}
+
%
{
{
\frac{1}{L}
\frac{c}{\pi}
\frac{1}{c^2 +  \Lambda^2}
\ \ %
}
}
\label{eq:HW-int-eq-repulsive}
\ee
with the restricting conditions
\be
n&=&\frac12 \int_{-Q}^Q \rho(k) \,\textrm{d}k - \frac{1}{2 L}, \quad 
n_\downarrow =\frac12\int_{-B}^B \sigma(\Lambda) \,\textrm{d}\Lambda  -  \frac{1}{2 L} 
\nonumber\\
e&=& \frac12 \int_{-Q}^Q k^2 \rho(k)\,\textrm{d}k.
\ee
This agrees with the result for the general $su(N)$ fermionic case~\cite{Fujimoto}, when the
boundary potentials in their model become infinitely strong and the extra terms vanish.
The same calculation for attractive interactions and hard walls leads to
\be
\rho(k)&=& \frac{1}{\pi} + \frac{2}{\pi}
\int_{-B}^B
\frac{c\, \sigma(\Lambda) \textrm{d}\Lambda }{c^2 + 4(k-\Lambda)^2}
-\frac{2}{\pi L}
\frac{c}{c^2 + 4 k^2}
\nonumber\\
\sigma(\Lambda)&=&
\frac{2}{\pi} + \frac{1}{\pi} \int_{-B}^B
\frac{c\, \sigma(\Lambda') \, \textrm{d}\Lambda' }{c^2 + (\Lambda-\Lambda')^2}
+
\frac{2}{\pi}  \int_{-Q}^Q
\frac{c\, \rho(k) \, \textrm{d}k }{c^2 + 4(\Lambda-k)^2}
\nonumber\\
&& \quad -\frac{1}{\pi L}  \frac{c}{c^2 +  \Lambda^2}
-\frac{2}{\pi L}  \frac{c}{c^2 + 4 \Lambda^2}
\label{eq:HW-int-eq-attractive}
\ee
with the restricting conditions
\be
\fl
\quad n &=& 
\frac12
\int_{-Q}^Q  \rho(k)\, \textrm{d}k
+
\int_{-B}^B
\sigma(\Lambda)\,\textrm{d}\Lambda
- \frac{3}{2 L},
\qquad
n_\downarrow-n_\uparrow
=\frac12 \int_{-Q}^Q \rho(k)\,\textrm{d}k
- 
\frac{1}{2 L}
\nonumber\\
\fl \quad e &=&
\frac12 \int_{-Q}^Q 
k^2 \rho(k)\, \textrm{d}k
+
\int_{-B}^B 
\left(
\Lambda^2 - {c^2}/{4}
 \right)
\sigma(\Lambda)\,\textrm{d}\Lambda.
\ee

In these equation terms proportional to $\frac1L$ are `surface energy' terms, 
which appear for the hard wall case compared to the periodic case.
The $L$ appearing after taking the thermodynamic limit is understand to be large but not infinite.
The only effect of the different boundaries appears for this still finite $L$.
Apart from these terms the integral equations for
the attractive and repulsive case are the same as those for the periodic case.
The above hard wall integral equations can be used to obtain
expressions for the surface energy and then compared to the results 
obtained by taking a continuum limit in the Hubbard model \cite{Schulz}.

Note that in the derivation of the ground state integral equations in the attractive regime 
bound pairs of approximate form $k_\alpha=\Lambda_\alpha \pm \frac12 \mathrm{i}{c}$
were assumed (see also eqn. 2.114 in Ref.~\cite{Takahashi}).
This is no contradiction to the dependency of the bound momenta on the coupling strength 
going as $\sqrt{c}$ in Section (\ref{sec:weak}).
The equations are valid in different regimes, for weak coupling we examine a finite size 
(i.e., $L$ and $N$ finite) with vanishingly small coupling constant $c$.
In contrast, in this section we consider a system at fixed finite $c$ and take the thermodynamic limit
$N,L\to\infty$ with $N/L=\textrm{const}$.
These limits do not commute, see the discussion in Ref.~\cite{ours-hw-bosons}

\section{Concluding remarks}
\label{sec:outlook}

In this paper we have extended the Bethe Ansatz solution for one-dimensional bosons and fermions
with $\delta$-interaction and arbitrary internal degrees of freedom to the case of hard wall boundary conditions.
The general solution \eref{eq:HW-1st-level-BAE-suN}
and \eref{eq:suN-HW-higher-levels} relies on the known Bethe Ansatz solution of the open $su(N)$ chain.
The special case of the two-state fermionic model has been analyzed in detail, 
both for periodic and hard wall boundary conditions.
Expansions were obtained for the ground state energy in the weak and strong coupling limits 
in both the attractive and repulsive regimes. 
Integral equations were also given for the ground state.
We expect that given the recent experimental advances on one-dimensional quantum
gases the exact solution to the family of models considered here will 
enjoy further interest, particularly for the investigation of finite systems with a boundary.

\ack
The authors thank Professor M. Takahashi for some helpful discussions.
Financial support from the Australian Research Council through the Discovery and Linkage International
programs and the Deutsche Forschungsgemeinschaft 
(grant BO2538/1-1) is gratefully acknowledged.

\clearpage

\appendix
\section{Coordinate Bethe Ansatz for $\delta$-interaction quantum gases}
\label{appa}
In this Appendix we give the coordinate Bethe Ansatz solution for the $\delta$-interaction gas 
with hard wall boundary conditions.
For ease of comparison we also discuss the more standard case of periodic boundary conditions.
Our aim is to give a self-contained treatment.
For further background the reader is referred to the 
books \cite{Korepin,Takahashi,Sutherland,Gaudinbook,Mabook} and 
where appropriate, to the original literature.
Like for the periodic case, the problem is solved by making contact with results for the
diagonalisation of the $su(N)$ chain, the solution of which is discussed in \ref{appb}.

Unlike the spin chains, the Hamiltonian does not have a mechanism for spin flipping.
However, the model is block diagonal for different spin arguments $\{N_1,\ldots,N_{N'}\}$,
where $N_i$ is the number of coordinates $\sigma_i$ having value $i$.
For example, for spin-$\frac12$ particles the number of spin $\uparrow$ and spin $\downarrow$ 
particles is conserved.
We want (anti-)symmetry under coordinate exchange, as we are considering indistinguishable particles.
The 2-particle $\delta$-interaction scatters elastically, i.e., at most momenta can be exchanged, 
but not their value (however, scattering at a reflecting wall might change $k_i\to -k_i$, as discussed below).
The contact interaction $\delta(x_i-x_j)$ only plays a role `near' the hyperplanes, 
where two coordinates are the same $x_i\to x_j$. 
In quantum mechanics books \cite{practical-QM,Mabook} it is shown by integrating over the region boundary, 
that the wave function is continuous but has a finite jump in the first derivative. 
Thus generally these problems can be seen as simple free particle problems away from the
hyperplanes $x_i\to x_j$, with plane wave solutions. 
These regions with free solutions are connected by appropriate boundary conditions, 
accommodating the jump in the 1st derivative from the contact interaction. 
This consideration leads to the Ansatz wave functions 
\be
\fl
\quad
\Psi(x_1,\ldots,x_N,\sigma_1,\ldots,\sigma_N) \nonumber \\
\fl 
\quad\quad
=\sum_{P}  A_{\sigma_1,\ldots,\sigma_N}(P_1,\dots,P_N|Q_1,\ldots,Q_N) 
\mathrm{e}^{\mathrm{i}(k_{P1}x_{Q1}+ \ldots +k_{PN}x_{QN})}
\label{eq:ansatz-wavefunction}
\ee
\be
\fl
\quad
\Psi(x_1,\ldots,x_N,\sigma_1,\ldots,\sigma_N)  \nonumber \\
\fl
\quad\quad
=\sum_{P}\!\! \sum_{\epsilon_{Pi}=\pm}
 A_{\sigma_1,\ldots,\sigma_N}(\epsilon_{P1},\ldots,\epsilon_{PN}|Q_1,\ldots,Q_N) 
\mathrm{e}^{\mathrm{i}(\epsilon_{P1}k_{P1}x_{Q1}+\ldots+\epsilon_{PN}k_{PN}x_{QN})}
\label{eq:ansatz-wavefunction-hw}
\ee
for arguments $x_i \neq x_j$ $\forall i \neq j$ with $i,j=1,\ldots,N$.
%
These wavefunctions are for periodic and hard wall boundary conditions, respectively.

In \eref{eq:ansatz-wavefunction} and \eref{eq:ansatz-wavefunction-hw}
the sums run over all $N!$  permutations $P$ of $\{1,2,\ldots,N\}$, 
and over all $N$ signs $\epsilon_i=\pm$.
The wavefunctions are piece-wise defined functions -- 
for each given argument $(x_1,\ldots,x_N)$ there is exactly one permutation $Q$, so that
$0\leq x_{Q1}<x_{Q2}<\ldots<x_{QN}\leq L$. 
For \eref{eq:ansatz-wavefunction} there are $N$ complex, pair-wise numbers $\{k_i\}$ and 
$N! \times N! \times N'^N$ complex numbers $A_{\sigma_1,\ldots,\sigma_N}(P|Q)$ labelled by the 
permutations $P, Q$ of the combinations and signs.
Similarly for \eref{eq:ansatz-wavefunction-hw} there are $N! \times N!\times N'^N \times 2^N$ complex numbers
$A_{\sigma_1,\ldots,\sigma_N}(\eps_{P1}...\eps_{PN}|Q)$.

We turn now to the key properties of the wavefunction.\\
%
{\em (Anti-)Symmetry.}
Applying the (anti-)symmetry conditions $\Psi\to\Psi$ (bosons) and $\Psi\to-\Psi$  for (fermions)
under particle exchange $(x_i,\sigma_i)\leftrightarrow(x_j,\sigma_j)$ to the wave functions
\eref{eq:ansatz-wavefunction} and \eref{eq:ansatz-wavefunction-hw}
leads to (For brevity these formula are for hard walls, the periodic case is recovered by 
setting $\eps_{Pi}\to P_i$.)
\be
\fl \qquad
A_{\,\, \sigma_i \,\, \sigma_j \,\,}(\eps_{P_1},\ldots, \eps_{P_N}| \,\, Q_a \,\, Q_b \,\,) 
 = \pm
A_{\,\, \sigma_j \,\, \sigma_i \,\,}(\eps_{P_1},\ldots, \eps_{P_N}| \,\, Q_b \,\, Q_a \,\,)
 \label{eq:anti-cond-hwc}
\ee
where we have introduced an obvious abbreviation in the notation.
Here $Q_a$ and $Q_b$ denote the positions of $x_i$ and $x_j$ for a given argument of $\Psi$ and
we assume $x_i<x_j$ without loss of generality.
In the same way define $i=P_a$ and $j=P_b$. 
Introduce the operator $T^{ij}$ via
$\left(T^{ij}\right)_{\sigma_1\ldots\sigma_N}^{\sigma'_1\ldots\sigma'_N}=
\pm\delta_{\sigma_i,\sigma'_j}\delta_{\sigma_j,\sigma'_i}\prod_{r\neq i,j}\delta_{\sigma_r,\sigma'_r}$, 
i.e., $T^{ij}={\cal P}^{ij}$ for bosons and $T^{ij}=-{\cal P}^{ij}$ for fermions, 
in terms of the permutation operator ${\cal P}^{ij}$.
The (anti-) symmetry condition can then be rewritten, using the summation
convention for internal matrix indices from now on, as
\be
{\left(T^{ij}\right)}_{\sigma_1\ldots\sigma_N}^{\sigma'_1\ldots\sigma'_N}
A_{\sigma'_1\ldots\sigma'_N}(\eps_{P_1},\ldots,\eps_{P_N}| \,\, Q_a \,\, Q_b \,\,) \nonumber\\
\qquad =A_{\sigma_1\ldots\sigma_N}(\eps_{P_1},\ldots,\eps_{P_N}| \,\, Q_b \,\, Q_a \,\,).
\label{eq-T-def}
\ee
{\em Continuity.} 
At the intersection of two regions the piece-wise definition of the wave function 
$\Psi(x_1,\ldots,x_N,\sigma_1,\ldots,\sigma_N)_{x_i<x_j \atop x_i\to x_j}=
\Psi(x_1,\ldots,x_N,\sigma_1,\ldots,\sigma_N)_{x_i>x_j \atop x_i\to x_j}$, 
requires that the coefficients satisfy
\be
\fl \qquad
A_{\sigma_1\ldots\sigma_N}(\,\,\eps_{P_a}\,\,\eps_{P_b}\,\,|\,\, Q_a\,\, Q_b\,\,)
 + 
A_{\sigma_1\ldots\sigma_N}(\,\,\eps_{P_b}\,\,\eps_{P_a}\,\,|\,\, Q_a\,\, Q_b\,\,)
\nonumber\\
\fl \qquad\qquad =
 A_{\sigma_1\ldots\sigma_N}(\,\,\eps_{P_a}\,\,\eps_{P_b}\,\,|\,\, Q_b\,\, Q_a\,\,)
 +
 A_{\sigma_1\ldots\sigma_N}(\,\,\eps_{P_b}\,\,\eps_{P_a}\,\,|\,\, Q_b\,\, Q_a\,\,).
\label{eq:cont-cond}
\ee
{\em $\delta$-interaction.} 
In standard QM textbooks (e.g., \cite{practical-QM,Mabook}), it is shown by integration
that in local center-of-mass coordinates $X, y$ for the two coordinates in the argument 
$x_i=X_{Qa}\to x_j=X_{Qb}$ the $\delta$-interaction leads to
\be
\left. \frac{\partial \Psi}{\partial y} \right|_{y=0^+}
-
\left. \frac{\partial \Psi}{\partial y} \right|_{y=0^-}
= c\, \Psi|_{y=0}.
\label{eq:jump-cond-3}
\ee

Applying this last result to the wavefunction results in a 
relation involving four $A$ coefficients with different permutations $P$ and $Q$, 
one of which can be eliminated by using \refeq{eq:cont-cond}.
In this way
\be
\fl \qquad
\mathrm{i} (k_{P_b} - k_{P_a})
\left[
A_{\,\,\sigma_i\,\,\sigma_j\,\,}(\,\, P_a \,\, P_b \,\, | \,\, Q_a \,\, Q_b \,\,)
- A_{\,\,\sigma_i\,\,\sigma_j\,\,}(\,\, P_b\,\, P_a\,\, | \,\, Q_b \,\, Q_a \,\,)
\right]
\nonumber\\
\fl \qquad\qquad
= c \left[
A_{\,\,\sigma_i\,\,\sigma_j\,\,}(\,\, P_a \,\, P_b \,\, | \,\, Q_a \,\, Q_b \,\,)
+ A_{\,\,\sigma_i\,\,\sigma_j\,\,}(\,\, P_b\,\, P_a\,\, | \,\, Q_a \,\, Q_b\,\,)
\right].
\ee

This equation, connecting $A$ coefficients with different permutations $P$ and $Q$ but
identical spin indices $\sigma_i$, can be transformed, using the operator $T$ from \refeq{eq-T-def}, 
into an equation connecting $A$ with a different permuation $Q$, but the same permutation $P$ 
and spin indices $\sigma_i$ by `hiding' the spin coordinate permutation inside the definition of $T^{ij}$.
The resulting central equation was found by Yang \cite{Yang}.
In our notation it takes the form of a matrix equation 
\be
\fl
A_{\sigma_1\ldots\sigma_N}(\,\, P_a \,\, P_b \,\,| \,\, Q_a \,\, Q_b \,\,)
= 
\underbrace{ 
\left[
\frac{\mathrm{i} (k_{P_b} - k_{P_a}) T^{ij} + c \, \textrm{Id}}
{\mathrm{i} (k_{P_b} \!-\! k_{P_a}) - c}
\right]_{\sigma_1 \ldots \sigma_N}^{\sigma'_1 \ldots \sigma'_N}
}
 A_{\sigma'_1\textrm{\tiny...}\sigma'_N}
(\,\, P_b \,\,P_a \,\, | \,\, Q_a \,\, Q_b \,\,)
 \nonumber
%
\ee
in the spin indices $\sigma_1\ldots\sigma_N$.
Here the highlighted term defines the operator $Y^{ij}(k_{P_b} - k_{P_a})$.
The formula for hard walls is obtained by replacing $k_i\to\eps_i k_i$ and 
also $P_i \to \eps_{Pi}$ in the coefficients. 
For future reference we introduce the operators
$\left[X^{ij}(u)\right]_{\sigma_1\ldots \sigma_N}^{\sigma'_1\ldots\sigma'_N}
:=\left[Y^{ij}(u)\right]_{\sigma''_1\ldots\sigma''_N}^{\sigma'_1\ldots\sigma'_N}
\left[T^{ij}(u)\right]^{\sigma''_1\ldots\sigma''_N}_{\sigma_1\ldots\sigma_N}$, 
which connect coefficients with permutations
$P$ and $Q$ differing by the same transposition only
\be
\fl
\quad
A_{\sigma'_1\ldots\sigma'_N}(\,\, P_a \,\, P_b \,\, | \,\, Q_a \,\, Q_b \,\, )
=\left[ X^{ij}(k_{Pb}-k_{Pa}) \right]_{\sigma_1 \ldots \sigma_N}^{\sigma'_1\ldots\sigma'_N}
A_{\sigma'_1\ldots\sigma'_N}(\,\, P_b \,\, P_a \,\, | \,\, Q_b \,\, Q_a \,\,)
\nonumber
\ee
So far restrictions have only been imposed on the coefficients $A$, while the $N$ complex
$k_i$ are arbitrary (the `scattering problem').
Imposing boundary conditions restricts the choice of these values to the so called Bethe equations, 
which we now turn to.

\subsection{Periodic Boundary Conditions}
Without losing generality the periodicity condition can be written as 
$\Psi|_{x_1=0}=\Psi|_{x_1=L}$, which when applied 
to the ansatz wave function \refeq{eq:ansatz-wavefunction} leads to the condition
\be
A_{\sigma_1\ldots\sigma_N}(P_1, P_2,\ldots,P_N |Q_1,\ldots,Q_N ) 
\nonumber\\
\qquad
= \mathrm{e}^{\mathrm{i} \, k_{P_1} L}
A_{\sigma_1\ldots\sigma_N}(P_2,\ldots,P_N, P_1 |Q_1,\ldots,Q_N ).
\label{eq:PBC-condition}
\ee
With the help of repeated application of the $X$ operators defined above this can be rewritten
as an equation connecting coefficients $A(P|Q)$ with identical permutations $P$ and $Q$ through 
a series of matrix multiplications.
The way of commuting $Q_1$ and $P_1$ through needs to be path-independent to be consistent. 
This is ensured by the Yang-Baxter equation, which is satisfied by the $X$ operators. 
This can be seen by the identification of the $R$ and $X$ operators in \refeq{eq:id-R-matrix-with-X} 
below and then using \refeq{eq:YBE}.

Applying successive multiple $X$ operators leads to the set of $N$ equations 
\be
\fl \qquad
A_{\sigma_1\ldots\sigma_N}(1... i-\!1,i+\!1... N,i\,|1... i-\!1,i+\!1...N,i) \,
 \mathrm{e}^{\mathrm{i} \, k_{i} L}\nonumber \\
\fl \qquad\qquad
=X^{1 i}(k_{1}-k_{i})_{\vec{\sigma}}^{\vec{\sigma}'''} \!\! \ldots  \,
X^{i-1 i}(k_{{i-1}}-k_{i})_{\vec{\sigma}^{iv}}^{\vec{\sigma}^{v}}
X^{1 i+1}(k_{{i+1}}-k_{i})_{\vec{\sigma}^{v}}^{\vec{\sigma}^{vi}} \ldots 
\nonumber
\\
\fl \qquad\qquad\quad
\dots X^{N i}(k_{N}-k_{i})_{\vec{\sigma}'}^{\vec{\sigma}''}
A_{\sigma''_1\ldots\sigma''_N}(1 ... i-\!1,i+\!1 ... N,i |
1 ... i-\!1,i+\!1 ... N,i)
\nonumber
\ee
which can be simultaneously diagonalized.
Here we have introduced the abbreviated notation $\vec{\sigma}=(\sigma_1...\sigma_N)$.
Note also that $P_2=1,\ldots,P_N=N$ and we choose $P_1=i$ and adapt the indices accordingly.
This means it is possible find linear combinations of $A_{\vec{\sigma}} (P|Q)$ 
considered as vectors in the indices $\sigma_1\ldots\sigma_N$.
From now on we suppress the permutations, as it is always the same argument $(P|Q)$ 
in the remaining equations.
Call these linear combinations say, $M_{\vec{\sigma}}^{\vec{\sigma}'} A_{\vec{\sigma'}}$, so that
the right hand side product of $X$ operators can be replaced with its eigenvalue 
$\theta(k_{i}|k_{1}\ldots k_{N})$
via $\mathrm{e}^{\mathrm{i}k_i L} M A = X \ldots X M A = \theta(\ldots) M A$, 
resulting in the 1st level Bethe equations
\be
\mathrm{e}^{\mathrm{i}\,k_{i} L}
=\theta(k_{i}|k_{1}\ldots k_{N})
\qquad \forall \ i=1,\ldots,N.
\label{eq:EV-theta}
\ee

The eigenvalue $\theta$ can be obtained on rewriting the product of $X$ operators in terms of 
$R$ matrices via the identification
\be
X^b(u)= \frac{u-\mathrm{i}\,c}{u+\mathrm{i}\,c}
R({\mathrm{i}u}/{c})
\qquad 
X^f(u)=R(-{\mathrm{i}\,u}/{c}).
\label{eq:id-R-matrix-with-X}
\ee
Note that the identification for bosonic and fermionic systems is different.
The $R$-matrix is defined in \refeq{eq:def-R-matrix}.
The $X$ operators contain a signed permutation operator for either type of particle, 
resulting from the condition of (anti-)symmetry under particle exchange (we suppress matrix indices).
Thus
\be
\fl \quad
X^{1 i}(k_{{1}}-k_{i}) \dots  
X^{i-1 i}(k_{{i-1}}-k_{i})
X^{1 i+1}(k_{{i+1}}-k_{i})
\ldots X^{N i}(k_{N}-k_{i})
\nonumber\\
\fl \qquad
= 
\left[
\prod_{j\neq i}^N
\frac{k_i -k_j+\mathrm{i}c}{k_i -k_j- \mathrm{i}c}
\right]
\textrm{tr}_0 \!\left[
R_{10}\left(-\frac{\mathrm{i}}{c}(u-k_1)\right) \ldots R_{N0}\left(-\frac{\mathrm{i}}{c}(u\!-\!k_N)\right)
\right]_{u={k_i} }
%
\nonumber\\
\fl \qquad
=
\textrm{tr}_0 \left[
R_{10}\left(\frac{\mathrm{i}}{c}(u-k_1)\right)\ldots R_{N0}\left(\frac{\mathrm{i}}{c}(u-k_N)\right)
\right]_{u={k_i} }.
%
%
\label{eq-R-product}
\ee
Here we have added a trace operation over an additional space 0 of the same type as the other $N$ spaces.
The second and third lines in this equation are for bosonic and fermionic $X$, respectively.

Because the $R$ matrix reduces to the permutation operator at 
$R_{i 0}({\mathrm{i}}(k_i-k_i)/c)=P^{i 0}$, 
the trace can be evaluated and this product reduces to a product of the $X$ operators.
The operator in the square bracket is the monodromy matrix of the periodic $su(N)$ chain.
Its eigenvalue was originally obtained by Sutherland \cite{Sutherland}. 
We give the resulting eigenvalue and also sketch the technically more interesting 
diagonalisation of the open $su(N)$ chain, of relevance to the hard wall $\delta$-interaction quantum gases, 
in \ref{appb}.

Using the eigenvalue $\Lambda(u)$ from equation \eref{eq:periodic-ABA-EV}
gives the 1st level Bethe equations (\ref{eq:EV-theta}) in the form
\be
\fl \qquad 
\mathrm{e}^{\mathrm{i} k_i L}= \prod_{j\neq i}^M \frac{k_i - k_j + \mathrm{i}c  }{  k_i - k_j - \mathrm{i}{c}  }
\,\,\, 
\Lambda \left( u=-\frac{\mathrm{i}}{c} k_i \right.\left| \{q_j\}=\{-\frac{\mathrm{i}}{c} k_j\}\right)
&\qquad \textrm{(bosons)}
\nonumber
\\
\fl \qquad
\mathrm{e}^{\mathrm{i}k_i L}=
 \Lambda\left( u=+\frac{\mathrm{i}}{c} k_i \right.\left| \{q_j\}=\{+\frac{\mathrm{i}}{c} k_j\}\right)
&\qquad \textrm{(fermions)}
\nonumber
\ee
for $i=1,\ldots,N$.
Using a shift in the roots gives the standard form \cite{Sutherland} 
\be
\fl \qquad
\mathrm{e}^{\mathrm{i}k_i L} =
\prod_{j\neq i}^N \frac{k_i - k_j + \mathrm{i}c  }{  k_i - k_j - \mathrm{i}{c}  }
\ \ \prod_{j=1}^M \frac{k_i - \Lambda_j -\frac12 \mathrm{i}{c} }{  k_i - \Lambda_j + \frac12 \mathrm{i}{c}  }
&\qquad \textrm{(bosons)}
\\
\label{eq:periodic-1st-level-BAE}
\fl \qquad
\mathrm{e}^{\mathrm{i}k_i L} =
\prod_{j=1}^M \frac{k_i - \Lambda_j + \frac12 \mathrm{i}{c}  }{  k_i - \Lambda_j - \frac12 \mathrm{i}{c}  }
& \qquad \textrm{(fermions)}
\ee

The momenta $k_1,\ldots,k_N$ are constricted by $M$ additional
Bethe equations (here for $su(2)$, the general $su(N)$ case is given in 
equation \eref{eq:suN-periodic-BAE})
\be
\prod_{j \neq i}^M \frac{\Lambda_i -\Lambda_j + \mathrm{i}c}{\Lambda_i -\Lambda_j - \mathrm{i}c}
=
\prod_{j =1}^N \frac{\Lambda_i -k_j + \frac12 \mathrm{i}{c}}{\Lambda_i -k_j - \frac12 \mathrm{i}{c}}
\qquad \forall\ i=1,\ldots,M.
\ee
The energy eigenvalue associated with the Bethe wavefunction is given in terms of the 1st level Bethe
roots, regardless of the number of nested levels, by
\be
E=\sum_{i=1}^N k_i^2
\nonumber
\ee
because the initial ansatz was just a combination of plane waves with appropriate boundary conditions.
To obtain the general solution for $N'$ internal states insert the general eigenvalue given in 
\refeq{eq:periodic-ABA-EV} into the right hand side of 
\eref{eq:EV-theta}
and use the $N'-1$ levels of Bethe equations in \refeq{eq:periodic-BAE-higher-level}.


\subsection{Hard Wall Boundary Conditions}
\label{sec:appa-HW}
The hard wall boundary condition $\Psi|_{x_i=0}=\Psi|_{x_i=L}=0$ for $i=1,\ldots,N$, 
translates into the equivalent of \refeq{eq:PBC-condition} when applied to the wave function 
\eref{eq:ansatz-wavefunction-hw}, namely
\be
A_{\sigma_1\ldots\sigma_N}(\eps_{P_1}\textrm{\small =+},\eps_{P_2},\ldots,\eps_{P_N}|Q_1, 
Q_2,\ldots,Q_N ) 
\nonumber\\
\qquad
= - \mathrm{e}^{\mathrm{i} k_{P_1} L  }  
 A_{\sigma_1\ldots \sigma_N}(\eps_{P_1}=-,\eps_{P_2}, \ldots, \eps_{P_N}|Q_1, 
Q_2,\dots,Q_N )
\nonumber\\
\mathrm{e}^{\mathrm{i} k_{P_1} L  }  
A_{\sigma_1\ldots\sigma_N}(\eps_{P_2},\ldots,\eps_{P_N},\eps_{P_1}=\textrm{\small +}|
Q_2,\ldots,Q_N, Q_1) 
\nonumber\\
\qquad
= - A_{\sigma_1\ldots\sigma_N}(\eps_{P_2},\ldots,\eps_{P_N}\eps_{P_1}=- |
Q_2,\ldots,Q_N, Q_1).
\nonumber
\ee
The strategy is similar to that for the periodic case described in the previous section.
With the help of the $X$ operators both equations can be reduced to contain the same permutations 
$P$ and $Q$ as arguments. 
Then one of the $A$ coefficients, say the one containing $\eps_{P_1}=+$, is eliminated,
leading to a single matrix equation \footnote{Again we set $P_1=i$, $P_2=1, \ldots, P_N=N$ 
and similarly for $Q$. Note that this time the term with $X^{ii}$ is omitted twice.} 
as in the periodic case:
\begin{eqnarray}
\fl
\nonumber
\qquad \mathrm{e}^{\mathrm{i}2 k_{i} L }
A_{\sigma_1\dots\sigma_N}(\eps_{i}=-,\eps_{1},\dots,\eps_{N}| i, 1, 2 \,\, N )
\nonumber\\
\fl
\qquad\quad=
\left[
X^{1 i}(\eps_{1}k_{1}-k_{i} ) \ldots 
X^{N i}(\eps_{N}k_{N} -k_{i} ) 
X^{N i}(-[\eps_{{N}} k_{{N}} + k_{i} ]) \ldots
\right.
\nonumber\\
\fl
\qquad\qquad
\ldots \left.
X^{1 i}(-[\eps_{1} k_{1} + k_{i}] )
\right]_{(\sigma_1\ldots\sigma_N)}^{(\sigma'_1\ldots\sigma'_N)}
%
%
%
A_{\sigma'_1\ldots\sigma'_N}(\eps_{i}=-,\eps_{1},\ldots,\eps_{N}|i, 1, 2,\ldots,N).
\nonumber
%
\end{eqnarray}

This reduction is again consistent, i.e., independent of the path of transpositions, 
due to the Yang-Baxter equations.
This expression can again considered to be an eigenvalue equation in which the product of
$X$ operators on the right hand side has the eigenvalue $\theta^{HW}(k_i|k_1\ldots k_N)$, 
leading to
\be
\mathrm{e}^{\mathrm{i} 2 k_{i} L} =\theta^{HW}(k_{i}|k_{1}\ldots k_{N})
\ee
Using the same idea as in the periodic case, this eigenvalue
can be obtained in terms of the eigenvalue of the open $su(N)$ chain, 
which is derived in \ref{appb}, at a special value of the argument. 
Taking the trace in the auxiliary space has to be done more carefully,
in order to account for the term $X^{i0}{(-2 k_i)}$, which does not leave a prefactor, because 
$\textrm{tr}_0\, R_{i0}(-\mathrm{i}{2k_i}/{c})= 1$.
We now set $\eps_n=+$ without loss of generality.
The next step is analogue to the periodic case, using the identification from \refeq{eq:id-R-matrix-with-X}
to obtain
\be
\fl
\nonumber
X^{1 i}(k_{1}-k_{i}) \ldots 
X^{i-1 i}(k_{i-1} -k_{i})
X^{i+\!1 i}(k_{i+1} - k_{i})
\ldots 
X^{N i}(k_{N} \!-\!k_{i}) \ldots
\nonumber\\
\fl
\ldots X^{N i}(-[ k_{{N}} + k_{i} ]) \ldots 
X^{i+\!1 i}(-[ k_{i+1} + k_{i}] )
X^{i-\!1 i}(-[ k_{i-1} + k_{i}] )
\ldots
X^{1 i}(-[ k_{1} + k_{i}] )
\nonumber\\
\fl
=\left[
\prod_{j\neq i}^N
\frac{c\!-\!\mathrm{i}( k_j \!-\!k_i)}{c\!+\!\mathrm{i}( k_j \!-\! k_i)}
\frac{c\!-\!\mathrm{i}( k_j \!+\!k_i)}{c\!+\!\mathrm{i}( k_j \!+\! k_i)}
\right]
\textrm{tr}_0 T_0\left(u=-\frac{\mathrm{i}}{c} k_i+\frac12 \right| \left. q_j=-\frac{\mathrm{i}}{c} k_j+\frac12   \right)
\nonumber\\
\fl
=
\textrm{tr}_0 T_0\left(u=\frac{\mathrm{i}}{c} k_i +\frac12 \right| \left. q_j=\frac{\mathrm{i}}{c} k_j+\frac12  \right).
%
\ee
The last two lines in this equation are for bosons and fermions, respectively.
Using the eigenvalue $\Lambda^{\textrm{\tiny HW}}(u)$ of $\textrm{tr}_0 T_0(u)$ calculated in 
\refeq{eq:HW-ABS-EV-suN}  the 1st level Bethe equation can be written as
\be
\fl
\mathrm{e}^{\mathrm{i}2 k_i L}=
\prod_{j=1}^M
\frac{k_i - \Lambda_j + \frac12 \mathrm{i}{c} }{k_i - \Lambda_j -\frac12 \mathrm{i}{c}}
\frac{k_i + \Lambda_j +\frac12 \mathrm{i}{c} }{k_i + \Lambda_j -\frac12 \mathrm{i}{c}}
%
&\quad
\textrm{(fermions)}
\nonumber\\	
\fl
\mathrm{e}^{\mathrm{i}2 k_i L}=
\prod_{j\neq i}^N
\frac{ k_i -k_j \!+\! \mathrm{i}c}{ k_i - k_j \!-\! \mathrm{i}c}
\frac{ k_i +k_j \!+\! \mathrm{i}c}{ k_i + k_j \!-\! \mathrm{i}c}
\,\,
\prod_{j=1}^M
\frac{k_i - \Lambda_j -\frac12 \mathrm{i}{c} }{k_i - \Lambda_j +\frac12 \mathrm{i}{c}}
\frac{k_i + \Lambda_j -\frac12 \mathrm{i}{c} }{k_i + \Lambda_j +\frac12 \mathrm{i}{c}}
&\quad
\textrm{(bosons)}
\ee

Again the 1st level Bethe quasi-momenta $\{ k_1,\ldots,k_N \}$ are
governed by a second set of parameters, the spin roots $\Lambda_1,\ldots,\Lambda_M$, 
which obey the Bethe equations \refeq{eq:HW-BA-su2-higer-levels}.
The most general $su(N)$ case is given in \eref{eq:suN-HW-higher-levels}, for $su(2)$ 
they reduce to
\be
\fl \qquad
\prod_{j\neq i}^M
 \frac{\Lambda_i - \Lambda_j + \mathrm{i}c	}{\Lambda_i - \Lambda_j -\mathrm{i}c}
\
 \frac{\Lambda_i + \Lambda_j	 +\mathrm{i}c	}{\Lambda_i + \Lambda_j - \mathrm{i}c}
\,\, \prod_{j=1}^N
%
%
 \frac{\Lambda_i - k_j -\frac12 \mathrm{i}{c}	}{\Lambda_i - k_j + \frac12 \mathrm{i}{c}	}
\
 \frac{\Lambda_i +k_j  - \frac12 \mathrm{i}{c}	}{\Lambda_i + k_j  + \frac12 \mathrm{i}{c} }
=1 
\ee
valid for both bosons and fermions. 
These equations are invariant under a simul\-taneous change of sign of all parameters within 
either set, $\{k_i \}$ or $\{\Lambda_i\}$.


\section{Algebraic Bethe Ansatz for the $su(N)$ chain}
\label{appb}
In this Appendix we give the necessary details for the diagonalization of the $su(N)$ chain, 
first for periodic and then for hard wall boundary conditions.
Our approach here is via the Algebraic Bethe Ansatz (see, e.g., Ref.~\cite{Korepin}), 
although we could have equally used the analytic Bethe Ansatz \cite{MN,YB}.

\subsection{Periodic boundary conditions}
 {
The diagonalization of the product of $R$-operators in equations (\ref{eq-R-product}) 
was first done in the context of the periodic $su(N)$ chain \cite{Sutherland}. 
In the Algebraic Bethe Ansatz formalism conventions and prefactors differ, often making comparison 
between different papers difficult.
Here we follow the normalization notation given by the rational $R$-matrix 
\be
\fl
\qquad
R_{ij}(u)=\frac{u \textrm{Id}_{ij} + {\cal P}_{ij}}{1+u}
\quad
\Leftrightarrow
\quad
R_{\alpha \beta}^{\gamma \delta}(u)=
\frac{u}{1+u} \delta_{\alpha,\gamma}\delta_{\beta,\delta}
+
\frac{1}{1+u} \delta_{\alpha,\delta}\delta_{\beta,\gamma}
\label{eq:def-R-matrix}
\ee
The standard notation for its entries are $a(u)=1$,
$b(u)={u}/{(1+u)}$ and $c(u)={1}/{(1+u)}$.
$R$ acts on the two $N'$-dimensional vector spaces $V_i$ and $V_j$,
${\cal P}$ is the permutation operator.
Spaces with indices $1,\ldots,N$ are denoted as quantum spaces, 
with index $0$ the auxiliary space.
All relations involving $R$-matrices are simple matrix multiplications with 
summations over internal indices. 
Proofs are most simply done in a graphical notation~\cite{deVega}, here
we only give the final key equations here.
The above $R$-matrix is normalized, with inverse (on both spaces)
$R_{ij}(-u)R_{ij}(u)={\rm Id}_{ij}$.
It satisfies the Yang-Baxter relation (YBE)
\be
R_{12}(u-v)R_{13}(u)R_{23}(v)=R_{23}(v)R_{13}(u)R_{12}(u-v).
\label{eq:YBE}
\ee
These equations are sufficient to prove all other relations in this section. 

For the periodic case the monodromy matrix has $N$ arbitrary inhomogeneities 
$\{q_1,\ldots,q_N\}$ and is given by
\be
T_0(u)= R_{10}(u-q_1) \ldots R_{N0}(u-q_N).
\nonumber
\ee
For these conventions the eigenvalue $\Lambda(u)$ of the transfer matrix, the $0$-trace
of the monodromy matrix $T_0(u)$, is given by
\be
\fl \qquad
\Lambda(\mu|\{q_i\})=&
\underbrace{
\prod_{j=1}^N a(u-q_j)
}_{=1}
&\prod_{j=1}^{M_1}
\frac{u-\mu^{(1)}-1  }{ u-\mu^{(1)} }
+ \textrm{vanishing terms}
\label{eq:periodic-ABA-EV}.
\ee
Here {\sl vanishing terms} denotes the remaining part of the periodic $su(N)$ chain eigen\-value,
which vanishes due to the special argument $u=q_j$ under consideration for the quantum gas model.
The product over the $N$ roots is equal to unity due to the choice of $R$-matrix normalization.
In the complete eigenvalue expression all functions containing terms of the form $\prod_{j=1}^N b(u-q_i)$ 
vanish and only the term stemming from the action of $T_{1}^{1}(u)$ in the monodromy matrix 
remains (for the full expression see Refs.~\cite{Sutherland,Korepin}).
Generally it is most convenient to obtain the Bethe equations
\be
\fl \qquad
&&
\prod_{j=1}^{N}
\frac{ \mu_k^{(1)} - q_j +1}{\mu_k^{(1)} - q_j}
=
\prod_{j=1 \atop j\neq k}^{M_{1}}
\frac{ \mu_k^{(1)}-\mu_j^{(1)}+1}{\mu_k^{(1)} -\mu_j^{(1)} -1}
\prod_{j=1}^{M_{2}}
\frac{\mu_k^{(1)} - \mu_j^{(2)}-1}{\mu_k^{(1)} - \mu_j^{(2)}}
\nonumber\\
\fl \qquad
&&
\prod_{j=1}^{M_{l-1}}
\frac{\mu_k^{(l)}-\mu_j^{(l-1)}+1}{\mu_k^{(l)}-\mu_j^{(l-1)}}
=
\prod_{j=1 \atop j\neq k}^{M_{l}}
\frac{\mu_k^{(l)}-\mu_j^{(l)}+1}{\mu_k^{(l)}-\mu_j^{(l)}-1}
%
\prod_{j=1}^{M_{l+1}}
\frac{\mu_k^{(l)}-\mu_j^{(l+1)}-1}{\mu_k^{(l)}-\mu_j^{(l+1)}}
%
%
\nonumber
\\
\fl \qquad
&&
\prod_{j=1}^{M_{N'-2}}
\frac{\mu_k^{(N'-1)} - \mu_j^{(N'-2)}+1}{\mu_k^{(N'-1)}-\mu_j^{(N'-2)}}
=
\prod_{j=1 \atop j\neq k}^{M_{N'-1}}
\!\!
\frac{\mu_k^{(N'-1)} - \mu_j^{(N'-1)} + 1}{
	\mu_k^{(N'-1)} - \mu_j^{(N'-1)} - 1}
\label{eq:periodic-BAE-higher-level}.
\ee
from the full expression of the eigenvalue as a pole-free condition than
finding the Bethe equations from the requirement that `unwanted terms'
in the Algebraic Bethe Ansatz cancel. 
Here $k=1,\ldots,M_l$ in each line, with $l=2,\ldots,N'-2$ in line two.

The standard form of the Bethe equations for the quantum gas model 
has a more symmetric form obtained by a shift in the parameters 
$\mu^{(k)}_i \to \mp {\mathrm{i}}\Lambda^{(k)}_i{c^{-1}} - \frac12{k}$,
in which the upper (lower) sign is for bosons (fermions) 
(cf. equation \eref{eq:suN-periodic-BAE}).
Details on performing the calculations above can be found in
Refs.~\cite{Sutherland,Sutherlandbook,Korepin,Gaudinbook} among others.

\subsection{Hard Wall boundary conditions}

In this section the relevant diagonalization of the double-row transfer matrix was first done 
for the $su(2)$ case \cite{Sklyanin} in the context of the open $XXZ$ spin chain.
A pedagogical derivation of the general $su(N)$ case via the Algebraic Bethe Ansatz 
has also been given \cite{deVega}.
The quantum gas model requires only the special limit of the rational $R$-matrix, with
all reflection matrices being simply scalar multiples of the identity matrix. 
We will state the results and key steps adapted from Ref.~\cite{deVega}
to the special case at hand.

First define the monodromy matrix $T_0(u)$ and its transfer matrix $\tau(u)$ with
$N$ arbitrary inhomogeneities $q_i$
\be
\fl \qquad
T_0(u)=\frac{2u+1 }{2u+2} R_{10}(u+q_1) \ldots R_{N 0}(u+q_N) R_{N 0}(u-q_N) 
\ldots R_{1 0}(u-q_1)
\nonumber
\\
\fl \qquad \tau(u)=\frac{2u+1 }{2u+2} \, \textrm{tr}_0 T_0(u)
\label{eq:def-transfermatrix}
\ee
acting on the quantum and auxiliary spaces, and the quantum spaces, respectively.
In addition to the YBE one has the `reflection equation' 
\be
\fl
&
R_{12}(u-v) T_1(u) R_{12}(u+v) T_2 (v)&=
T_2 (v) R_{12}(u+v) T_1(u) R_{12}(u-v)
\nonumber
\\
\fl
\Leftrightarrow\quad
&
R^{kl}_{\gamma\eta}(u-v) T^\gamma_\alpha(u) R_{i\beta}^{\alpha\eta}(u+v) T_j^\beta (v)
&=
T_\eta^l (v) R^{k \eta}_{\gamma \beta}(u+v) T^{\gamma}_{\alpha}(u) R^{\alpha \beta}_{ij}(u-v).
\label{eq:monodromy-commutation-generator}
\ee
The commutativity of the transfer matrices $[\tau(u),\tau(v)]=0$ can be verified by 
taking the trace of the reflection equation over the auxiliary space.
When applying the Algebraic Bethe Ansatz it is convenient
to consider the monodromy matrix as a matrix in auxiliary space,
with entries acting as operators on the quantum spaces, namely 
\be
T_0(u)=\left({
%
\begin{tabular}{cccc}
 $T_{1}^1(u)$ & $T_{2}^1(u)$ & $\ldots$ & $T_{N'}^1(u)$ \\
$T^2_{1}(u)$ &$T^2_{2}(u)$   & $\ldots$  &$T^2_{ N'}(u)$ \\
$\vdots$ & $\vdots$ &   $\ddots$  &  $\vdots$\\
$T^{N'}_1(u)$ & $T^{N'}_2(u)$ & $\ldots$   & $T_{N'}^{N'}(u)$  
\end{tabular}
}\right)
\ee
Define the vacuum state 
$|\textrm{VAC}\rangle=|1\rangle_{V1}\otimes \ldots \otimes |1\rangle_{VN}$,
the state in which all sites are in the same state
(for convenience we label the internal states by $1,\ldots,N'$).
The action of the monodromy matrix $T_0(v)$ on this vacuum state is 
\be
T(u) | \textrm{VAC} \rangle
=
\left({
%
%
\begin{tabular}{cccc}
 $A(u) | \textrm{VAC} \rangle$ & 
 $| {\textrm{\tiny some}\atop \textrm{\tiny state}} \rangle$ & $\ldots$ & 
 $| {\textrm{\tiny some}\atop \textrm{\tiny state}} \rangle$ \\
$0$ & $B(u) | \textrm{VAC} \rangle $   & $\ldots$  & $\vdots$ \\
$\vdots$ & $\vdots$ &   $\ddots$  &  $\vdots$\\
$0$ & $0$ & $\ldots$   & $B(u) | \textrm{VAC} \rangle$ 
\end{tabular}
}\right)
\nonumber
\ee
with the scalar functions $A(u)$ and $B(u)$ evaluated below. 
Note that entries are zero, except for the main diagonal ($T_{i}^{i}$ has the vacuum state as eigenvector,
with vacuum eigenvalues $A(u)$ and $B(u)$) and the operators $T_{j}^1$ on the first line,
which create some unknown state. 
The entries in the first line are creation operators $B_j(u):=T_{j}^1(u)$.
The reflection equation leads to more complicated commutation relations\footnote{
Obtained by taking special values of the external indices and evaluating the known values of the 
$R$-matrix in 
\refeq{eq:monodromy-commutation-generator}.
E.g., $(ijkl)=(2211),(2111),(2221)$ for the three commutation relations given.}
between elements of the monodromy matrix compared to the periodic case.

For notational simplicity we restrict the calculation from here on to the special case
$N'=2$, corresponding to particles with at most two internal spin states.
In this case there is only one creation operator, $B(u):=B_2(u)$.
Reading off the general commutation relations for arbitrary $N'$ leads 
to a nested form \cite{Sutherland,Sklyanin,deVega} of the Bethe Ansatz.
Here $\left[  B(u),B(v)  \right] =0$ with 
\be
\fl \qquad
T_1^1(v) 	B(u ) &=&
\underbrace{
\frac{a(u-v) b(u+v)}{b(u-v) a(u+v)} 	B(u )	T_1^1(v)
}
\nonumber\\
\fl
&& - \frac{c(u-v)	b(u+v)}{b(u-v)	a(u+v)}	B(v) 	T_1^1(u)
-
\frac{1}{a(u+v)}	B(v) 	T_2^2(u)
\\
\fl \qquad
T_2^2(v )	B(u) &=&
\underbrace{  \frac{u-v+1}{u-v} \frac{u+v+2}{u+v+1}}
B(u) 	T_2^2(v) 	
\nonumber\\
\fl \qquad\qquad
&&+
\frac{u+v+2}{(u-v)(u+v+1)} 
B(v)  T_2^2(u)		
+
{\frac{u-v-2}{(u-v)(u+v+1)}}
B(v) T_1^1(u)
\nonumber\\
\fl \qquad\qquad
&&+
%
\underbrace{\frac{2}{(u-v)(u+v+1)}}
B(u) T_1^1(v).
%
%
\ee

The Algebraic Bethe Ansatz provides a systematic procedure for obtaining a complete set of
eigenstates and eigenvectors of the transfer matrix $\tau(u)=T_{1}^{1}(u)+T_{2}^{2}(u)$.
As there is only one creation operator $B(v)$ for the $su(2)$ case,
the ansatz for the Bethe state is
\be
|\Phi\rangle = \prod_{i=1}^M B(\mu_i) |\textrm{VAC}\rangle
\ee
which is shown to be an eigenstate of $\tau(u)$ for certain complex numbers $\{\mu_1,\ldots,\mu_M\}$
and all arguments $u$.
The strategy is to commute the operators $T_{i}^{i}(u)$ of the main diagonal through to the right 
of the creation operators $B(\mu_i)$ to act directly on the vacuum state.
As only terms proportional to $|\Phi\rangle$ can be part of the eigenvector,
only terms {\em not} exchanging the arguments (highlighted in the commutation relations above) 
can contribute to the eigenvector -- the `wanted terms'.
All other terms, arising from the remaining terms in the commutation relation in any 
commutation step, are `unwanted terms' which can be shown to vanish for certain 
$\{\mu_1,\ldots,\mu_M\}$. 
These conditions are the Bethe equations.
In the periodic case only one term in the commutation relation is `wanted', making it possible to 
collect all wanted terms from the commutation operations in one product.
Thus a simple linear transform $\tilde{T}_{2}^{2}(u)=(2 u+1)T_2^2(u) -T_1^1(u)$ 
is applied to obtain the commutation relations in the same familiar form
\be
\fl \qquad
T_1^1(v) B(u) &=&
\underbrace{
{
\frac{a(u - v)	b(u+v)}{b(u - v)	a(u+v)}  B(u )	T_1^1(v)}}_{\textrm{wanted term}}
+
\frac{b(2u)}{a(2u) b(v - u)}
B(v) 	T_1^1(u)
\nonumber\\
&&-
\frac{1}{a(u + v) a(2u)}B(v) 	\tilde{T}_2^2(u)
\ee
\be 
\fl \qquad
\tilde{T}_2^2(v) B(u)&=&
\underbrace{
{
\frac{a(v-u)a(u+v+1)}
{b(v-u)a(u+v)}
B(u) \tilde{T}_2^2(v)
}}_\textrm{wanted term}
+
\frac{a(2v+1)}{a(2u)b(u-v)}
B(v) \tilde{T}_2^2(u)
\nonumber\\
\fl
&&+
\frac{a(2v+1)b(2u)}{a(u+v)a(2u)}
B(v) {T}_1^1 (u).
\ee

Combining everything together the transfer matrix eigenvalue $\Lambda(u)$ is given by
\be
\tau(u) |\Phi \rangle
&=&
\frac{2u+1}{2u+2}
\left[\frac{2u+2}{2u+1}T_1^1 (u) + \frac{1}{2u+1}\tilde{T}_2^2(u) 
\right]|\Phi \rangle
\nonumber\\
&=&\underbrace{
\left[
\Lambda_1^1 (u) 
+ 
\frac{1}{2u+2}\tilde{\Lambda}_2^2(u) 
\right]}_{\Lambda(u)} 
|\Phi \rangle
\ee
with the `wanted' contributions from the diagonal elements
\be
\Lambda_1^1(u)&=&
\underbrace{
\prod_{i=1}^N
a(u-q_i) a(u+q_i)
}_{\textrm{vacuum eigenvalue}}
\,
\underbrace{
\prod_{i=1}^M \frac{a(\mu_i-u)	b(\mu_i+u)}{b(\mu_i-u) a(\mu_i+u)} }_{\textrm{commutation relation}}
\\
%
\tilde{\Lambda}_2^2(u)&=&
Q(u)
\underbrace{\prod_{i=1}^M 
\frac{a(u-\mu_i)	}{b(u-\mu_i)	} 
\frac{	a(u+\mu_i+1)}{	a(u+\mu_i)} 
}_{\textrm{commutation relation}}.
%
%
\ee
The vacuum eigenvalue $Q(u)$ of $\tilde{T}_2^2(u)$ has the required simple form
if calculated as
\be
\fl \qquad
\tilde{T}_2^2(u)| \textrm{VAC}\rangle
=
\left[(2u+1) T_2^2(u) - \prod_{j=1}^N a(u+q_i)a(u-q_i) \right]
|\textrm{VAC}\rangle.
\ee
The vacuum eigenvalue $Q_2(u)$ of $T_2^2(u)$ is obtained via 
additional commutation relations, using the YBE at $v=-u$, resulting in
\be
Q_2(u)&=&
\frac{c(2u)}{a(2u)}
\left[
\prod_{j=1}^N a(u+q_i)a(u-q_i) -\prod_{j=1}^N
b(u+q_i)b(u-q_i)
\right]
\nonumber\\
&&+\prod_{j=1}^N  b(u+q_i) b(u-q_i).
\ee

The final expression for the eigenvalue is given by
\be
\Lambda(u)&=&
\frac{2u	}{2u +2 }
\prod_{i=1}^M
%
\frac{u-\mu_i+1}{u-\mu_i }
\frac{u+\mu_i+2}{u+\mu_i+1} \,
%
\prod_{i=1}^N
\frac{u+q_i}{u+q_i +1}
\frac{u-q_i}{u-q_i +1}
%
\nonumber\\
&&+
\prod_{i=1}^M
\frac{u-\mu_i-1}{u+\mu_i +1}
\frac{u+\mu_i}{u-\mu_i }.
\label{eq:HW-ABS-EV}
\ee
In principle it needs to be shown that the unwanted terms vanish and that the conditions for this are
the Bethe equations.
For simplicity we content ourselves here with reading 
off the conditions which make the eigenvalue $\Lambda(u)$ pole-free at the parameters $\mu_j$, 
yielding the Bethe equations in the form (with $i=1,\ldots,M$)
\be
\fl \qquad
 \prod_{j= 1}^M
 \frac{\mu_i - \mu_j -1	}{\mu_i - \mu_j +1	}
 \prod_{j\neq i}^M
 \frac{\mu_i + \mu_j		}{\mu_i + \mu_j +2	}
%
%
\prod_{j=1}^N
 \frac{\mu_i +q_j +1	}{\mu_i + q_j 	}
\ 
 \frac{\mu_i - q_j	+1	}{\mu_i - q_j 	}
%
%
=-1.
\ee
Applying the shifts $\mu_k\to \mathrm{i}\Lambda_k c^{-1}-\frac12$ and 
$q_l\to \mathrm{i} k_l c^{-1}$ gives the form normally
used in the $\delta$-interaction literature, namely
\be
 \fl \qquad
\prod_{j \neq i}^M
 \frac{\Lambda_i - \Lambda_j + \mathrm{i}c	}{\Lambda_i - \Lambda_j -\mathrm{i}c	}
\
 \frac{\Lambda_i + \Lambda_j	 +\mathrm{i}c	}{\Lambda_i + \Lambda_j - \mathrm{i}c	}
%
%
\prod_{j=1}^N
 \frac{\Lambda_i -k_j  - \frac12 \mathrm{i} c	}{\Lambda_i - k_j  + \frac12 \mathrm{i} c	}
\ 
 \frac{\Lambda_i + k_j	- \frac12 \mathrm{i} c}{\Lambda_i + k_j + \frac12 \mathrm{i} c	}
%
%
=1.
 \label{eq:HW-BA-su2-level2}
\ee

The Bethe equations \refeq{eq:HW-BA-su2-level2}
are the condition for \refeq{eq:HW-ABS-EV}
to be the eigenvalue of the transfer matrix $\tau(u)$ defined in  \refeq{eq:def-transfermatrix}.
Note that in order to make contact with the eigenvalue formulas in \ref{sec:appa-HW}
one needs to apply the shift $u\to u-\frac12$ in the above formula.

The eigenvalue $\Lambda^{\textrm{\tiny HW}}(u)$ for the general $su(N)$ case can be obtained
in similar fashion, with 
\be
\fl
\Lambda^{\textrm{\tiny HW}}(u)=
\label{eq:HW-ABS-EV-suN}
%
%
%
%
%
%
\prod_{j=1}^{M_0}
\frac{u + \mu^{(0)}_j  -1 }{u+ \mu^{(0)}_j}
\frac{u - \mu^{(0)}_j   }{u- \mu^{(0)}_j+1}
%
\left[
\sum_{k=1}^{N'}
\frac{2u-1}{2u+k-2}
\frac{2u}{2u+k-1}
\right.
\nonumber
\\
\fl
\times
\left.
\prod_{j=1}^{\!M_{k-1}}
\frac{u+\!\mu_j^{(k-1)} +k - 1  }{  u+\mu_j^{(k-1)} +k -2   }
\frac{u-\mu_j^{(k-1)}  +1  }{  u-\mu_j^{(k-1)}    }
\prod_{j=1}^{M_{k}}
\frac{u+\mu_j^{(k)} + k - 2  }{  u+\mu_j^{(k)} + k -1   }
\frac{u-\mu_j^{(k)}  - 1  }{  u-\mu_j^{(k)}    }
\right].
\ee
This formula is adapted from Ref.~\cite{deVega}, with 
$\mu_i^{(k)}\to \mp \mathrm{i} \Lambda_j^{(k)}c^{-1} - ({k-1})/{2}$,
with the upper (lower) sign for bosons (fermions) and the 
monodromy inhomogeneities $q_j=\mu^{(0)}_j\!+\!\frac12$.
Again all higher terms in the eigenvalue (i.e., sum terms with $k>1$) 
vanish for the special argument $u=\mu^{(0)}_j$ and $u=\mu^{(0)}_j+1$, 
due to terms of the form $\prod_{j=1}^N (u\textendash\mu^{(0)}_j)(u +\mu^{(0)}_j+1)$.
Here the conditions for the unwanted terms to cancel,  or equivalently, for the
eigenvalue to be pole-free, are given by
\be
\fl
\prod_{j=1}^{M_{k-1}  }
\frac{
\Lambda^{(k)}_i
-
\Lambda^{(k-1)}_j
+
\frac12  \mathrm{i}{c}
}{
\Lambda^{(k)}_i
-
\Lambda^{(k-1)}_j
-
\frac12  \mathrm{i}{c}
}
\
\frac{
\Lambda^{(k)}_i
+
\Lambda^{(k-1)}_j
+
\frac12  \mathrm{i}{c}
}{
\Lambda^{(k)}_i
+
\Lambda^{(k-1)}_j
-
\frac12  \mathrm{i}{c}
}
=
\prod_{j\neq i}^{M_{k}  }
\frac{
\Lambda^{(k)}_i
-
\Lambda^{(k)}_j
+
 \mathrm{i}c
}{
\Lambda^{(k)}_i
-
\Lambda^{(k)}_j
-
 \mathrm{i}c
}
\
\frac{
\Lambda^{(k)}_i
+
\Lambda^{(k)}_j
+
 \mathrm{i}c
}{
\Lambda^{(k)}_i
+
\Lambda^{(k)}_j
-
 \mathrm{i}c
}
\nonumber\\
\times
\prod_{j=1}^{M_{k+1}  }
\frac{
\Lambda^{(k)}_i
-
\Lambda^{(k+1)}_j
-
\frac12  \mathrm{i}{c}
}{
\Lambda^{(k)}_i
-
\Lambda^{(k+1)}_j
+
\frac12  \mathrm{i}{c}
}
\
\frac{
\Lambda^{(k)}_i
+
\Lambda^{(k+1)}_j
-
\frac12  \mathrm{i}{c}
}{
\Lambda^{(k)}_i
+
\Lambda^{(k+1)}_j
+
\frac12  \mathrm{i}{c}
}
\label{eq:HW-BA-su2-higer-levels}
\ee
with $k=1,\ldots,N'-1$ and $i=1,\ldots,M_k$ in the $k$-th equation.
Here the first and last Bethe equations follow 
from this compact form on setting  $M_0=N$ and $M_{N'}=0$,
as there are no Bethe roots $\Lambda^{(N')}_j$ of level $N'$.

\end{document}